\documentstyle[12pt]{article}
\begin{document}

\baselineskip 15.2pt
\parindent=1cm
\parskip 3mm

\title{Thermal charm production by massive gluons and quarks}

\author{
P\'eter L\'evai$^{1,2}$
 \  and  \ Ramona Vogt$^{2,3,4,\star}$ \\ \\
 ${}^{1}$KFKI Research Institute for Particle and Nuclear Physics, \\
POB. 49, Budapest, 1525, Hungary \\
 ${}^{2}$Institute for Nuclear Theory, University of Washington, \\
Seattle, WA 98195 \\
 ${}^{3}$Nuclear Science Division, Lawrence Berkeley National Laboratory, \\
Berkeley, CA 94720 \\
 ${}^{4}$Department of Physics, UC Davis, Davis, CA 95616 }

\footnotetext{$^\star$ This work was supported in part by the Director,
Office of Energy Research, Division of Nuclear Physics of the Office of High
Energy and Nuclear Physics of the U. S. Department of Energy under Contract
Number DE-AC03-76SF0098.}

\maketitle

\begin{abstract}
We investigate charm production in an equilibrated quark-gluon plasma
produced in heavy-ion collisions at RHIC and LHC. 
Effective quark and gluon masses are introduced
from thermal QCD calculations.  
Assuming a Bjorken-type longitudinal expansion and including
the influence of temperature dependent masses on the expansion,
we determine the total number of $c \overline c$
pairs produced in the quark-gluon plasma phase.
We calculate the
charm production rate at leading order with massive gluons  
and quarks and compare our result to charm production by
massless partons. 
We consider two different scenarios for the
initial conditions, a parton gas with a rather long kinetic equilibration time
and a minijet gas with a short equilibration time.  In a parton gas,
assuming $m_c=1.2$ GeV, we obtain a 
substantial enhancement over the thermal charm rate
from massless quarks and gluons, up to 4.9 secondary
charm quark pairs in Au+Au collisions at RHIC and 245 charm pairs in Pb+Pb 
collisions at the LHC.
\end{abstract}

\begin{center}
LBNL-39984 \ \ 
{\it Submitted to Phys. Rev. C.}
\end{center}

\newpage
\section{INTRODUCTION}

Two decades ago, studies of quantum chromodynamics (QCD) predicted
that quarks and gluons can be liberated from hadrons and, at
appropriately high energy densities, a new state of the matter, the
quark-gluon plasma (QGP) will be formed \cite{QGP}. Lattice-QCD calculations
have shown that the critical energy density of this phase transition
is \mbox{$\varepsilon_c \approx 2-3$ GeV/fm$^3$}. The critical temperature
is $T_c\approx 260$ MeV for a pure gluon plasma 
while $T_c \approx 150$ MeV when dynamical light quarks are included 
\cite{LatQCD}. 
These conditions are expected to be  produced
in ultrarelativistic nucleus-nucleus collisions.  Recent experiments have not 
unambiguously proved QGP formation at the CERN SPS where
$\sqrt{s} =$ 17-20 A GeV,
but in heavy-ion collisions at RHIC ($\sqrt{s}=200$ A GeV) and LHC
($\sqrt{s}=$5.5 A TeV) the required energy densities should be produced.

At ultrarelativistic 
energies, nucleus-nucleus collisions can be described by parton
interactions in perturbative QCD inspired models \cite{HIJING,PC}.
In this framework, hard and semi-hard scatterings among partons dominate 
the reaction dynamics. These collisions are able to drive the parton system 
toward thermal equilibrium 
very effectively \cite{QGPtime,BDMTW,EskoParton}, but
chemical equilibrium may not be established \cite{BDMTW,EskoParton}.
Equilibration strongly depends on the actual parton
densities in the colliding nuclei and the microscopic
processes on the parton level. 
The behavior of the parton densities at small $x$ is not 
fully established: newer parametrizations of the parton densities already
greatly increase the energy density of the initial state \cite{EskoParton}
and the introduction of softer semi-hard scatterings (a smaller $p_T$ cutoff
on the minijet cross section) could yield a parton system very close to 
equilibrium a short time after the primary collisions.

During the time-evolution of the parton system,
in parallel with the thermalization, many collective phenomena
can appear. The massless color
degrees of freedom, the quarks and gluons, are strongly interacting,
as described in the framework of finite
temperature QCD \cite{TFT}. In this theory,
temperature dependent thermal quark and gluon masses are introduced 
which can dramatically affect the phenomenological description of 
the time evolution. Massive fermionic and bosonic excitations
can be derived in many ways, {\it e.g.} by considering leading order 
medium effects on the QCD self-energy \cite{selfglue}.
Since the behavior of the QGP and its high-temperature excitations
are not well understood, it is challenging
to find measurable tests of this behavior in heavy ion collisions.

In general, open charm 
\cite{WaMu,hpoc,Openchrv1,Openchrv2,LMW}, direct photons
and dilepton production \cite{Photlep} can be used as direct probes of
the early parton dynamics and the evolution of the QGP.
Primary charm production from the initial nucleon-nucleon collisions 
has been calculated and can be substantial \cite{WaMu},
especially when up-to-date parameterizations of
the parton densities which increase at low $x$, are used 
\cite{hpoc,Openchrv2}.  
In the most recent calculation of
initial charm production \cite{Openchrv2}
in Au+Au collisions, extrapolated from $pp$ collisions,
a total of 9 (450) {\it initial} $c \overline c$ pairs
were produced at RHIC (LHC), primarily due to the
behavior of the parton densities at low $x$.  As shown in
ref.\ \cite{Openchrv2}, shadowing can reduce
the yield in the central region by up to a factor of two, depending
on the $c \overline c$ pair mass and transverse momentum.

The amount of
secondary charm production in the QGP phase is an open question
\cite{Openchrv1,Openchrv2,LMW,JR}.
If the parton system is dilute \cite{BDMTW,LMW}, secondary charm
production would be suppressed because of the low parton fugacities.
Furthermore, at RHIC the charm mass is 2-3 times larger than
the expected temperature scale, $T \sim 0.35 - 0.55$ GeV, 
and much greater than the bare light quark masses.  Thus in this case,
even in a fully equilibrated plasma charm production is not
significantly enhanced.
At the higher LHC energy, the predicted initial temperatures,
$T\approx 0.7-1$ GeV, are close to the charm mass. 
One can then expect a
larger thermal charm production rate than at lower energies, perhaps comparable
to the initial yield.

Charm production in the plasma by massless quarks and gluons is small
because the massless partons must be very energetic to overcome
the $c \overline c$ pair mass threshold.
However, the effective thermal quark and gluon masses generated by the
plasma could significantly enhance the thermal charm yield,
perhaps even approaching the initial yield after shadowing has been included.  
This additional charm production
is a possible probe of massive excitations in the QGP which 
could test the level of thermalization and
the evolution of the quark gluon plasma.
Furthermore a large charm quark multiplicity would
favor the production of exotic phenomena such as
multi-charm hadron production \cite{LPJZMul}.

The influence of massive gluons on strangeness production
has already been investigated \cite{BLM} and
found to have a negligible influence on the total strangeness production rate.
However, that calculation considered massless light quarks and a 
temperature region, $T\approx 200-250$ MeV, only slightly
above $T_c$.
Here we perform a similar calculation but with 
massive quarks as well as massive gluons and with the higher initial
temperatures
obtained from more recent estimates of the initial conditions
\cite{EskoParton,XKSW}.  Under these circumstances, the
thermal quark and gluon masses can generate a significant increase 
in secondary charm production.

In this paper we investigate the thermal
charm production rate in a fully
equilibrated QGP with massive quark and gluon degrees of freedom.
We compare our results with those from zero mass quarks and gluons. We will
assume both a pure gluon gas and a quark-gluon plasma.
We also distinguish between parton gas and minijet gas initial conditions.
We obtain the total number of
$c \overline c$ pairs produced during the lifetime of the plasma by
assuming a Bjorken-type longitudinal expansion for Au+Au collisions at RHIC
and Pb+Pb collisions at LHC.
The influence of temperature dependent masses on the expansion and
the speed of sound is also considered in detail.

\section{\bf EFFECTIVE QUARK AND GLUON MASSES}

In a strongly-interacting QGP, collective fermionic and bosonic
excitations  appear, as also seen in QED. These quasi-particles can be
interpreted as {\it massive quarks} and {\it massive gluons}. The 
propagators are modified in the medium because the color charges are 
dressed by
the interaction with their neighbors in the plasma phase.
The effective masses are related to the properties of the QGP and can thus 
be  characterized by their temperature and density dependence.
Using the high temperature approximation \cite{KlimWel} or,
equivalently, the hard thermal loop approximation \cite{BraPis}
which ensures a gauge invariant treatment, one can obtain the following
effective  equilibrium masses at zero chemical potential,
$\mu=0$, for quarks and gluons massless at $T=0$ \cite{TFT}:
\begin{eqnarray}
m^2_{q, {\rm th}} (T) &=& { {g^2 T^2} \over 6}  \\
m^2_{g, {\rm th}} (T) &=& { {g^2 T^2} \over 3}  (1 + {N_f \over 6} ) \ .
\label{mgluon}
\end{eqnarray}
We assume that the equilibrated QGP contains light $u$, $d$, and $s$ 
quarks and anti-quarks so that $N_f = 3$ and 
$m_{g, {\rm th}} = g T / \sqrt{2}$.
Since the strange quark has a non-zero bare mass, we consider an
approximate effective mass, ${\hat m}_s$, where
\begin{equation}
{\hat m}_s = \sqrt{m^2_{s,0} + m^2_{q, {\rm th}}} \ ,
\end{equation}
and $m_{s,0}= 150$ MeV. We neglect the small current mass of the light quarks
so that ${\hat m}_{q}\approx m_{q, {\rm th}}$.
The effective mass of the gluon is similar, ${\hat m}_{g}=m_{g, {\rm th}}$.
Charm quarks are assumed to be produced with their bare mass and dressed
later.  The hadronization of charm quarks does not influence the charm
production rate.

The temperature-dependent running coupling constant is
\begin{eqnarray} 
{g^2} \left(\frac{T}{T_c} \right) = 
\frac{24\pi^2}{(33-2N_f) \ln
[(19T_c/\Lambda_{\rm \overline{MS}})(T/T_c)]} \, \, . 
\label{g2t}
\end{eqnarray} 
where in SU(3) gauge theory, $T_c/\Lambda_{\rm \overline{MS}} = 1.78 \pm 0.03$
\cite{KMS}. With this dependence, the thermal gluon and quark masses 
increase approximately linearly with
in the temperature, $m_{g, {\rm th}} \approx (130 \ T/T_c + 36)$ MeV and
\mbox{$m_{q, {\rm th}} \approx m_{g, {\rm th}}/\sqrt{3}$,} where 
$T_c=150$ MeV.  Then at $T \approx 500$ MeV, ${\hat m}_g\approx 470$ MeV while 
at the lower temperature assumed earlier \cite{BLM}, $T \approx
300$ MeV,  ${\hat m}_g\approx 300$ MeV.
In ref. \cite{BLM,Kampf} massive gluons were introduced by reproducing
lattice-QCD energy density
and pressure results \cite{LQCD2}.
The effective gluon mass, assumed constant, was $M_g \approx 450-500$ 
MeV for $1.2 < T/T_c < 2.4$, $30-40$\% larger than our value in the
same region.

\section{\bf  CHARM QUARK PRODUCTION}

The processes relevant for $c \overline c$ pair production in the quark-gluon
plasma are the following:
\begin{eqnarray}
q + {\overline q} & \longrightarrow & c + {\overline c} \\
s + {\overline s} & \longrightarrow & c + {\overline c} \\
g + g  & \longrightarrow & c + {\overline c} 
\end{eqnarray}
An analogy can be drawn between charm and strangeness production by the plasma.
In early calculations of strangeness production, processes (5) and (7)
were treated at leading order with bare quark and gluon masses \cite{Strzero}.
On the other hand,
effective quark masses were used as infrared cut-offs 
simultaneously removing electric and magnetic infrared singularities, see
{\it e.g.} ref.\ \cite{QGPtime}.
Effective gluon masses were included in calculations of chemical
equilibration \cite{BDMTW} and strangeness production \cite{BLM} in the QGP, 
and in calculations of radiative energy loss in a
parton gas \cite{GyXNW94}. In these cases 
the bare Green's functions were used with the 
effective gluon mass but the quarks were not dressed.
Here we combine these approximations, using the bare
Green's functions containing dressed quarks and gluons and
calculating the production rates in
perturbative QCD at leading order.

In the calculation of strangeness production with massive gluons \cite{BLM}
one additional process was included, the decay of massive gluons. From
the thermal gluon mass, eq.\ (\ref{mgluon}), and the 
temperature dependent running coupling constant, eq.\ (\ref{g2t}), we estimate 
that the threshold for $g \rightarrow c \overline c$,
${\hat m}_g > 2 m_c$, will be reached only at $T \approx 20 \ T_c$.
We therefore neglect this channel.  We can also assume that the annihilation
of produced $c \overline c$ pairs is negligible.
The total charm production rate is thus:
\begin{equation}
{{dN_{c{\overline c}}}\over {d\tau}} = \left[ A_{g}(T) + 2 A_{q}(T) +
A_s(T) \right] \cdot V(\tau)
\label{chtot}
\end{equation}
where
\begin{equation}
A_{i}(T) = {1\over{n_i!}}
\int ds \int 2 \sigma_i(s) \sqrt{s (s-4 {\hat m}_i^2)} \ {{f_i(E_1)}\over {2
E_1}} {{f_{\overline i}(E_2)}\over {2 E_2}}
 {{d^3 p_1}\over {(2\pi)^3}} {{d^3 p_2}\over {(2\pi)^3}} 
\end{equation}
and $n_i$ is the number of identical particles in the initial state.
We calculate $A_q$ and $A_s$ separately since ${\hat m}_q \neq {\hat m}_s$. 
The Bose distribution is used for gluons, 
$f_g(E) = (e^{E/T} - 1 )^{-1}$,
and the Fermi distribution is used for quarks, $f_{q, \overline q}(E) = 
(e^{E/T} + 1)^{-1}$.
We consider a symmetric QGP with zero baryon and strangeness 
chemical potential, 
$\mu_q=\mu_s=0$.  The total production cross section for each channel,
$\sigma_i$, is calculated at center-of-mass energy $\sqrt{s}$.

The gluon-gluon fusion and quark-antiquark annihilation
rates can be written in the Lorentz-invariant form
\begin{eqnarray}
A_{i}(T) & = & {1\over {n_i !}} 
\int ds \int {{d^4 p_1}\over{(2 \pi)^3}} \int {{d^4 p_2}\over{(2 \pi)^3}}
\ 2 \ \sigma{(s)} \ \sqrt{s (s-4 m_i^2)} \ \delta(s-(p_1+p_2)^2)    
\nonumber \\
 && \mbox{} \cdot \delta(p_1^2 - m_i^2) \delta(p_2^2 - m_i^2) f_i(E_1)
f_i(E_2) \ , 
\end{eqnarray}
with four momenta $p_1 = (E_1, {\overrightarrow {{\bf p_1} }})$ and
$p_2 = (E_2, {\overrightarrow {{\bf p_2} }})$.
Following ref.\ \cite{BLM} this integral can be rewritten as
\begin{equation}
A_{i}(T) = {{1}\over{32 \pi^4}} {1\over {n_i !}} \int^{\infty}_{4 m_i^2} ds
\ \sigma_i(s)\  s\  (s - 4 m_i^2) \sum^{\infty}_{l=2}\  \sum^{l-2}_{k=-l+2}
(\pm 1)^l {{K_1(a_{kl})}\over {a_{kl}}} \ ,
\end{equation}
where the sum over index $k$ is incremented by 2, with the $+$ sign used 
for bosons and the $-$ sign for fermions. The
modified imaginary Bessel function, $K_1(a_{kl})$, has the argument
\begin{equation}
a_{kl} = { {\sqrt{s}}\over {2T}} \sqrt{ l^2 - \left( 1 -
{{4 m_i^2}\over s} \right) k^2 }  
\end{equation}
The use of a simple Boltzmann distribution 
instead of the Bose and Fermi distributions
is equivalent to reducing the sum to only the $l=2, k=0$ term with
$a_{0 2} = \sqrt{s}/T$. At high temperatures and high effective quark
and gluon masses, the Boltzmann distribution is a good approximation.  However,
here we will use the full sum.
\medskip

Once the time evolution of the volume, $V(\tau)$,
and the temperature, $T(\tau)$, are determined, we can calculate the
appropriate quark and gluon masses and the total charm pair production rate.

\subsection{\bf  ELEMENTARY CROSS SECTIONS}

Here we explain the calculation of the total $c \overline c$ subprocess 
production cross sections for massive quarks and gluons.  As discussed earlier,
we neglect $c \overline c$ pair annihilation as well as 
the decay of massive gluons.

When the light quarks have an effective mass, the square of the matrix element
for processes (4) and (5) is
\begin{equation}
\vert M_i \vert^2 = 2 g^4 
\frac{{\rm Tr}[\gamma^\mu (p\!\!\!/-{\hat m}_i) \gamma^\nu 
(q\!\!\!/ +{\hat m_i}) ]~
{\rm Tr}[\gamma_\mu (p_1\!\!\!\!/-m_c) \gamma_\nu (p_2\!\!\!\!/ + 
m_c)]}{(s - {\hat m}_g^2 + \Gamma^2/4)^2 + \Gamma^2 {\hat m}_g^2 }\ ,\ \ \ 
\label{mqqb1}
\end{equation}
where $p^\mu$ and $q^\mu$ are the four-momenta of the incoming quarks, 
$i=u$, $d$ and $s$,
and $p_1^\mu$ and $p_2^\mu$ are the four-momenta of the outgoing $c$
and $\overline c$ quarks.  The propagator has been modified by 
the finite mass and
width of the gluons as in ref. \cite{BLM}.
After evaluation of the traces, the square of the 
matrix element can be expressed as 
\begin{equation}
\vert M_{i} \vert^2 = d_i^2 \pi^2 \alpha_S^2 \ \frac{64}{9} \,
\frac{(m_c^2+{\hat m}_i^2-t)^2 + (m_c^2+{\hat m}_i^2-u)^2 + 
(2 m_c^2+2 {\hat m}_i^2)s}{(s - {\hat m}_g^2 + \Gamma^2/4)^2 + \Gamma^2 
{\hat m}_g^2 }\ ,\ \ \ \label{mqqb2}
\end{equation}
where $g^2 = 4\pi \alpha_S$ and $d_i=6$ for the spin and color degrees 
of freedom for each flavor.  
Note that eq.\ (\ref{mqqb1}) does not include the usual
average over the initial spin and color while eq.\ (\ref{mqqb2}) does include
these factors.  

The total cross section is obtained from the $t$ integration of 
$\vert M_{i} \vert^2|$,
\begin{equation}
\sigma_i(s) = {1\over {16 \pi s(s-4{\hat m}_i^2)}} \int^{t_+}_{t_-} dt \ 
\vert M_{i} \vert^2  
\end{equation}
with the limits
\begin{equation}
t_{\pm} = - \left( { {\sqrt{s-4{\hat m}_i^2}}\over 2} \mp  { {\sqrt{s-4
m_c^2}}\over 2} \right)^2  
\end{equation}
where the sum of the Mandelstam invariants $s$, $t$ and $u$ is
\begin{equation}
s + t + u = 2 {\hat m}_i^2 + 2 m_c^2 \, \, .
\end{equation}

For gluon fusion, (7), the total cross section can be obtained 
from the integral
\begin{equation}
\sigma_g(s) = {1\over {16 \pi s (s-4{\hat m}_g^2)}} \int^{t_+}_{t_-} dt \ 
\vert M_s + M_u + M_t \vert^2  
\end{equation}
with the integration limits
\begin{equation}
t_{\pm} = - \left( { {\sqrt{s-4 {\hat m}_g^2}}\over 2} \mp  { {\sqrt{s-4
m_c^2}}\over 2} \right)^2  \, \, .
\end{equation}
Detailed calculations of the invariant matrix elements, $M_s$, $M_u$ and 
$M_t$, are given 
in ref.\ \cite{BLM} for $s \overline s$ production by
transverse gluons of constant mass.  
The same expressions can be used here with
the change $m_s \rightarrow m_c$.  The sum of the Mandelstam 
invariants is now
\begin{equation}
s + t + u = 2 {\hat m}_g^2 + 2 m_c^2  \, \, . 
\end{equation}

We use two different values 
for the charm quark mass: $m_c=1.2$ GeV and $m_c=1.5$ GeV.  The lower value 
was found to produce agreement between charm production calculated to NLO
\cite{hpoc} and $pp$ total cross section data.  This value has also been
used in recent estimates of the charm contribution to the dilepton yield
\cite{Openchrv2}.  The larger value is somewhat more standard 
\cite{Openchrv1,WaMu,LMW} and, at energies 
near the charm production threshold, 
allows an all-order resummation of soft and virtual
gluon corrections \cite{sv2}.  For smaller charm quark
masses, the series cannot be resummed.

\subsection{THE HYDRODYNAMICAL MODEL}

We now discuss our calculations of the equation of state and the time evolution
of the plasma.  The pressure of an ideal gas of massive particles is
\begin{equation}
P = \sum_i \frac{g_i}{6 \pi^2} \int_0^\infty \frac{dk \, k^4}{\sqrt{k^2 +
{\hat m}_i^2}} [\exp{(\beta \sqrt{k^2 + {\hat m}_i^2})} \mp 1]^{-1} \, \, ,
\label{Peq}
\end{equation}
where $i=g, u, {\overline u}, d, {\overline d}, s, {\overline s}$. 
In the case of a symmetric plasma, $\mu_u=\mu_d=\mu_s = 0$, and we have 
$P_i=P_{\overline i}$. 
Since ${\hat m}_i$ is temperature dependent, the energy density, $\epsilon =
T dP/dT - P$, has an additional term proportional to the mass gradient
\begin{equation}
\epsilon = \sum_i \frac{g_i}{2 \pi^2} \int_0^\infty \frac{dk \, 
k^2}{\sqrt{k^2 + {\hat m}_i^2}} [\exp{(\beta \sqrt{k^2 + {\hat m_i}^2})} 
\mp 1]^{-1} \left(k^2 + {\hat m}_i^2 - {\hat m}_i T \frac{d{\hat m}_i}{dT} 
\right)\, \, .
\end{equation}

We assume a simple longitudinal expansion and follow the time evolution through
the entropy 
\begin{equation}
s(\tau) = s(\tau_0) (\tau_0/\tau) \, \, .
\end{equation}
We also calculate the square of the sound speed, defined as
\begin{equation}
c_s^2 = \frac{dP/dT}{d\epsilon/dT}
\, \, ,
\end{equation}
where $dP/dT = s = (\epsilon + P)/T$
as a check on how closely the evolution follows that of an ideal massless gas
with $c_s^2 = 1/3$.  The temperature gradient of the energy density is
\begin{eqnarray}
\frac{d\epsilon}{dT} & = & \sum_i \frac{1}{T} \frac{g_i}{2 \pi^2} 
\int_0^\infty 
\frac{dk}{\sqrt{k^2 + {\hat m}_i^2}} [\exp{(\beta \sqrt{k^2 + {\hat m_i}^2})} 
\mp 1]^{-1} \left\{ (k^2 + {\hat m}_i^2)^2 \right.  \nonumber \\
&  & \mbox{} \left. + \, \, (k^2 + {\hat m}_i^2) \left(3k^2 - 
2{\hat m}_i T \frac{d{\hat 
m}_i}{dT} \right) \right. \\
&  & \mbox{} \left. + \, \, T^2 \left[ ({\hat m}_i \frac{d{\hat m}_i}{dT})^2 - 
k^2 \left( \left( \frac{d{\hat m}_i}{dT}\right)^2 + {\hat m}_i \frac{d^2{\hat 
m}_i}{dT^2} + 2\frac{{\hat m}_i}{T}\frac{d{\hat m}_i}{dT} \right) \right]
\right\} \, \, . \nonumber
\end{eqnarray}
Note that the above expressions reduce to $c_s^2 = 1/3$ when ${\hat m}_i$ is
constant. 

The mass gradients can be expressed rather simply when ${\hat m}_i = 
cg(T)T$ and the current quarks and gluons are massless.  In
this case we have
\begin{eqnarray}
\frac{d{\hat m}_i}{dT} & = & \frac{{\hat m}_i}{T} \left[ 1 - \frac{1}{2} \left(
\frac{{\hat m}_i}{cT} \right)^2 \right] \\
\frac{d^2{\hat m}_i}{dT^2} & = & -\frac{1}{2} \frac{{\hat m}_i^3}{c^2T^4} 
\left[ 1 - \frac{3}{2} \left(
\frac{{\hat m}_i}{cT} \right)^2 \right]  \, \, .
\end{eqnarray} 
For the strange quark, ${\hat m}_s = \sqrt{m_{s,0}^2
+ m_q^2(T)}$, the derivatives are
\begin{eqnarray}
\frac{d{\hat m}_s}{dT} & = & \frac{1}{{\hat m}_s}\frac{dm_q}{dT} \\
\frac{d^2{\hat m}_s}{dT^2} & = & -\frac{1}{{\hat m}_s^3} \left( \frac{dm_q}{dT}
\right)^2 + \frac{1}{{\hat m}_s}\frac{d^2m_q}{dT^2} \, \, ,
\end{eqnarray} 
where $dm_q/dT = d{\hat m}_i/dT$ as above.  Including effective quark
and gluon masses tends to slow the evolution of the system as well as increase
the $c \overline c$ rate.

We need to fix the space-time volume to calculate the absolute number of 
produced charm pairs.  The particle number is obtained from 
\begin{equation}
N = \int n^\mu d \sigma_\mu = \rho \pi R^2 \tau \int_{-\eta^*}^{\eta^*} d\eta
\cosh \eta = \rho V
\end{equation}
where $V = 2\pi R^2 \tau \sinh \eta^*$.  To determine the maximum space-time
extent of the plasma in rapidity as a function of $\tau$, we use the total
available energy as a rough estimate:
\begin{eqnarray}
E_{\rm tot} & = & 2AE_{\rm beam} = \int T^{0 \mu} d \sigma_\mu \\ \nonumber
            & = & \pi R^2 \tau \int_{-\eta^*}^{\eta^*} d \eta [\epsilon
                   \cosh^2 \eta + P \sinh^2 \eta] \, \, .
\end{eqnarray}
{} From a comparison of the calculated $E_{\rm tot}$ with the available energy,
$2AE_{\rm beam}$, we can determine the value of $\eta^*$ and calculate 
the volume.
With this volume, the charm yield is calculated from eq.\ (\ref{chtot}).

\section{TIME EVOLUTION OF THE SYSTEM}

We will consider the hydrodynamical evolution of a
fully equilibrated plasma with two different sets of
initial conditions, $T_0$ and $\tau_0$, for RHIC and LHC collisions.
  
The first parameter set is based on the  {\bf parton gas} model
derived from the HIJING Monte Carlo code \cite{HIJING}.
The kinetic equilibration time, relatively long, is reached
when the momentum distributions are locally isotropic due to elastic
scatterings and the expansion of the system, $\tau_0 \sim 
0.5-0.7$ fm.  We use $T_0 = 550$ MeV and
$\tau_0 = 0.7$ fm at RHIC and $T_0=820$ MeV, $\tau_0 = 0.5$ fm at
LHC, as in the ideal case described in ref.\ \cite{XKSW}.  
With this model we do not distinguish between a gluon gas and 
a quark-gluon plasma in the estimate of $T_0$.

The second set of initial conditions was determined from estimates of minijet
production \cite{EskoParton}.  In this case, the momentum scale of the minijet
calculation sets the initial time, $\tau_0 \sim 1/p_T\leq 1/p_0$, and the
minijet yield determines the initial temperature.
For a typical value of the momentum scale, $p_0 \sim 2$ GeV, $\tau_0 
\sim 0.1$ fm.  Because the minijet yield
depends on the composition of the system, the initial temperature depends on 
the partonic degrees of freedom.  Since  minijets predominantly produce
gluons, if we consider only a gluon gas, the highest $T_0$ is obtained,
$T_0 = 445$ MeV at RHIC and 1140 MeV at LHC.  When light quark production is
included, the number density increases but $T_0$ decreases to 360 MeV 
at RHIC and 900 MeV at LHC.  The plasma resulting from the early equilibration
time and high temperature is referred to
hereafter as a {\bf minijet gas}.  

We compare and contrast the evolution of the plasma with the parton gas and 
the minijet gas in Au+Au collisions at RHIC in fig.\ 1 and in Pb+Pb
collisions at the LHC in fig.\ 2.  In each case, we show results for a pure
gluon system and a quark-gluon system both with massless and massive partons.
The time evolution of the temperature, energy density, square of the sound
speed in the medium and the volume of the plasma are given in each case.

The parton gas has a longer lifetime due to the longer equilibration time.  The
temperature evolution is shown in figs.\ 1(a) and 2(a) for the parton gas and
in 1(b) and 2(b) for the minijets.  The difference between the temperature
evolution with gluons alone and for the quark-gluon gas is seen to be small,
independent of whether or not the partons are massive.  The evolution slows
when the parton masses are finite.  We have cut
off the evolution at $T_c=150$ MeV.
The parton gas, with its slower evolution, remains above $T_c$ for $\tau \leq
10$ fm.  At RHIC energies the temperature of the minijet gas
drops below $T_c$ at $\tau \approx 1.5$ fm in the quark-gluon system.  The
pure gluon system, with its larger $T_0$, remains above $T_c$ for at least
2.5 fm at RHIC.  For both the parton and minijet gas,
the higher initial temperatures
at the LHC make the finite parton masses more effective in slowing the
evolution, in part because the effective parton masses are 
larger for the higher temperature.

The increase in the number of degrees of freedom between a pure 
gluon gas and a
quark-gluon gas is clearly reflected in the difference between the energy 
densities of the two systems, shown in figs.\ 1(c-d) and 2(c-d).  
Note that in both cases at RHIC, the energy density of the
massive gluon gas is actually reduced relative to the massless gluon gas.  This
is due to the temperature gradient of the gluon mass.  A similar effect is
also observable in the minijet quark-gluon gas.
In all these cases, the temperature is not large compared to the effective
masses, causing the reduction.  At the much higher temperatures of the LHC,
although the effective masses are increased, the energy density is 
increased by the inclusion of the finite parton masses.  We note that 
at RHIC, the energy density of the minijet gas actually drops
below 1 GeV/fm$^3$ for $\tau \geq 1$ fm, even though the temperature remains
above $T_c$ until $\tau \approx 1.5$ fm in the quark-gluon system and until
$\tau \approx 2.5$ fm in the gluon gas.  
This suggests that the assumption of an equilibrated minijet gas at
RHIC is perhaps questionable.

The speed of sound in the medium remains close to that of an ideal gas, as 
shown in figs.\ 1(e-f) and 2(e-f).  
As expected, in the massless gluon case, $c_s^2 \equiv
1/3$.  The deviation of the dot-dashed curve 
from the ideal gas result is due to the finite current strange
quark mass even though the light quarks and gluons are massless.  
In this case, the system moves further from the ideal gas behavior
at later times
as the temperature becomes comparable to $m_s$.  When the system is composed
of massive gluons only, the deviation from an ideal gas is largest because
the effective gluon mass is larger than the effective quark masses.  The
addition of the lighter massive quarks into the system tends to bring
the sound speed closer to the ideal gas value.  At the LHC, the 
higher temperature keeps the system closer to the ideal gas limit than at
RHIC.

The volume of the system, crucial to the determination of the charm yield,
increases as shown in figs.\ 1(g-h) and 2(g-h).  
Since the volume depends on the rapidity
extent of the plasma, the relatively lower energy density and pressure of the
gluon gas require a larger spatial extent to ensure that
the energy of the system is equal to $E_{\rm tot}$.  Because we
have changed the initial conditions according to the composition of the minijet
gas, the volume changes less than the parton gas volume.

\section{TOTAL CHARM YIELD}

In this section, we present our results 
on the $c \overline c$ production rates.
The parton gas produces the largest $c \overline c$ yield because of the
slower time evolution demonstrated in the previous section.
Since the time evolution of the quark and gluon effective masses determines
the relative enhancement of charm production by the massive excitations, we
show the time dependence of the effective masses in a quark-gluon system
in fig.\ 3. 
The finite current strange quark mass at zero temperature
results in a slightly higher effective strange quark mass compared to the 
effective light
quark mass.  It also reduces the time dependence of the strange quark effective
mass.  The larger slope of the gluon effective mass as a function of
temperature results in a faster decrease in gluon mass as a function of time.
We have shown the effective masses for as long as $T>T_c$ --
the minijet gas at RHIC
is at $T_c$ when $\tau \sim 1.5$ fm.
At later times, the finite masses become less effective for
producing charm.   We remark 
that the running coupling constant, $g$, is also a function of time.
The weaker coupling at later times also reduces the yield.  

The initial rate is approximately a factor of two larger when the
quark degrees of freedom are included.  At the beginning of the evolution,
the rate is nearly independent of the initial parton mass.  However, since
the system cools more slowly with massive initial partons, the rate is larger
at later times in the massive case. 
The number of $c \overline c$ pairs produced during the lifetime of the plasma
is found by multiplying the rate, eq.\ (11), by the volume from eq.\ (30).  
In most cases, charm
production only occurs during the early part of the evolution.  At RHIC,
production by the parton gas is essentially over after $\sim 3$ fm while
production from the minijet gas is ended by $\sim 0.5$ fm.  Note however, that
$c \overline c$ pairs continue to be produced at much later times 
when the partons are massive, especially at the LHC.  
This is particularly true for the massive quark-gluon gas.

The charm production rates and the number of produced $c \overline c$ pairs as
a function of time at RHIC are given in fig.\ 4 for $m_c = 1.2$ GeV and 
fig.\ 5 for $m_c = 1.5$
GeV.  The final charm pair yield in both cases is given in Table 1.
The parton gas results are given in fig.\ 4(a-b) and 5(a-b).
Here, although the production rate is larger when both 
quarks and gluons are
included, the final number of $c \overline c$ pairs produced during the
evolution of the system does not strongly depend on the composition of the
plasma at RHIC because the larger rate is compensated by a
correspondingly smaller volume.  The yield from a massless quark-gluon gas
is about 15\% larger than that from a massless 
gluon gas.  For massive initial quarks and gluons, the difference
in the composition changes the yield by only 8\%, as seen in Table 1.
When $m_c = 1.2$ GeV, the yield is increased 44\% for
a massive gluon gas relative to 
a massless gluon gas and 32\% in the massive
quark-gluon system.  If $m_c = 1.5$ GeV, the enhancement due to the massive
partons is reduced to 36\% for the gluons alone and to 29\% for the quark-gluon
gas.  The yield at the lower mass is about 3.5 times larger than when $m_c =
1.5$ GeV.  Some of the enhancement can be accounted for by the slower
temperature evolution.  After the system has evolved for 10 fm, the temperature
of the massive gluon gas is approximately 10\% larger than the massless gas.
The difference is 8\% when the quarks are included.  Note
particularly that the 5 thermal $c \overline c$ pairs produced with $m_c = 1.2$
GeV is only a factor of two less than that expected from the initial production
at this energy \cite{Openchrv2}.

The yield is much smaller from a minijet gas due to the
shorter equilibration time and the lower initial temperature. (In the minijet
gas, $T_0$ is reduced
24\% for gluons alone and 53\% for a quark-gluon system relative to the initial
temperature of the parton gas.)  In the minijet gas,
the charm yield is reduced by
a factor of 60-70 (for gluons) and 250-300 (for quarks and gluons) compared to
the parton gas.  The thermal charm yield from the minijet gas is thus
negligible compared to the initial charm rate \cite{Openchrv2}.
The influence of the massive partons is also reduced for the minijet gas.  For
$m_c = 1.2$ GeV, the yield is only increased over the massless case
by 23\% in a gluon gas and 14\% in
a quark-gluon gas.  The influence of the charm mass on the yield is also
stronger for the minijet gas---the yield decreases by a factor of 4-5 with the
larger charm mass.

The enhancement due to the effective parton mass is more substantial at the 
LHC, as seen in figs.\ 6 and 7 and in Table 2.  
The massive quarks and gluons have a more significant effect on the temperature
evolution of the parton gas.  The temperature is 27\%
higher after 10 fm with massive gluons and 20\% higher
for massive quarks and gluons.
The enhancement of the yield is also larger:  when $m_c =1.2$ GeV, the yield
is 94\% larger in the gluon gas and 70\% higher in the quark-gluon gas.
Increasing the charm mass only reduces the total yield by a factor of 2.5, thus
the yield is less dependent on the charm mass at the higher temperature.
Note that here the charm yield from the massive quark-gluon system, 
250 pairs after 10 fm,
is similar to the initial nucleon-nucleon rate
\cite{Openchrv2}.

The minijet gas is more effective at producing charm at LHC than at RHIC.  
Although
the initial time remains short, $T_0$ is 40\% larger in the minijet gluon
gas and 10\% larger in the minijet quark-gluon gas than
in the parton gas.  Therefore the minijet charm yield is only a factor of two
to six smaller than the parton gas yield and the minijet thermal charm yield
also becomes a significant fraction of the initial production.  
Note also that the enhancement due
to the massive partons is largest in the minijet gluon
gas--a factor of 2.5 increase over that from a gas of massless gluons 
because the temperature of the massive gluon gas is
50\% larger.

In Ref.\ \cite{Openchrv2}, with an ideal, massless, quark-gluon plasma,
one thermal charm pair was found at RHIC and 26 charm pairs at LHC with $m_c =
1.2$ GeV.  In that 
work, the initial temperature was nearly the same as the parton gas $T_0$ used
here but the initial time was closer to that of the minijet gas, $\tau_0 \sim
(3T_0)^{-1}$.  However, a larger, constant, $g^2$ kept the yield from being 
reduced.  When initial conditions identical to those used here
are chosen for the calculation 
of Ref.\ \cite{Openchrv2}, the results are 
quite similar.  While the details of the expansion are somewhat
different, we have 
checked that our massless charm production cross
sections are in exact agreement with those in Ref.\ \cite{Openchrv2}.

To study the dependence of the enhancement 
on the charm pair mass, we also calculate
the thermal charm pair mass distributions for RHIC in fig.\ 8 and for
the LHC in fig.\ 9.  Generally, the thermal charm mass distributions are 
steeper than those
charm pairs produced in the initial nucleon-nucleon interactions.  (See ref.\
\cite{Openchrv2} for the initial charm pair
mass distributions.)  The shapes of the
mass distributions for each of the four cases we have studied with our two sets
of initial conditions are quite similar, especially for the parton
gas where $T_0$ is the same in all cases. The shapes of the
distributions from the minijet gas at RHIC are also similar although
the initial yield is larger from the gluon gas because of its higher $T_0$.
The enhancement is generally 
largest for low mass charm pairs, not far above threshold.
At LHC, where the enhancement is greatest, the distributions with the massive
excitations included approach the mass distributions for the massless case only
at $M \sim 8-10$ GeV.

\section{DISCUSSION}

We have investigated a new mechanism for enhancing thermal charm production
by a quark-gluon plasma: massive excitations in the plasma state.
We chose two different sets of initial conditions, a parton gas and a minijet
gas, and calculated the thermal charm yield from each for both massless and
massive quarks and gluons.
In our calculation we assumed that the system stayed in thermal and
chemical equilibrium during the expansion.  Therefore, our charm yield is 
an upper bound on secondary
charm production.   

The largest charm production was found when a parton gas was assumed because
the characteristic thermalization time was $\tau_0=0.5-0.7$ fm.
Then the parton gas lifetime was long, more then 10 fm.
At RHIC we obtained a 30-40\% enhancement
with massive gluons and quarks while at the LHC, a 50-100\% enhancement may be
expected.  With $m_c=1.2$ GeV, the charm mass used in recent calculations,
we obtained 4.9 secondary
charm quark pairs at RHIC and 245 charm pairs at LHC.
Note that these numbers, upper limits on secondary charm production, are
similar to the expected initial charm production \cite{Openchrv2}.
If we take $m_c=1.5$ GeV, secondary charm pair
production is reduced to 1.3 pairs at RHIC and 94 pairs at LHC.

The lifetime of the minijet gas was significantly shorter than the parton gas
due to the short thermalization time, $\tau \sim 0.1$ fm.  
This difference strongly
reduced the charm yield from the minijet gas compared to the parton gas.
The typical lifetime of the minijet gas was of order 2.5 fm 
although at RHIC the temperature of the
quark-gluon gas dropped below $T_c$ after only 1.5 fm.  The fast expansion
reduces the influence of the massive quasi-particles at RHIC 
where the initial temperature of the minijet gas
was also smaller than the initial temperature of the parton gas.
The enhancement was typically 20\% with massive quarks and gluons but
the total yield, 0.016 pairs from the quark-gluon gas, 
was very small compared to the initial charm rate.
Although the expansion was also fast at the LHC, the much higher initial
temperature, $T_0\approx 1$ GeV, generated large charm production rates.
With $m_c=1.2$ GeV we obtained 38 secondary
charm quark pairs from a quark-gluon gas and 100 pairs from a gluon gas while
with $m_c = 1.5$ GeV, the yield was 16 and 47 pairs respectively.  Note that
the minijet gluon gas result is only a factor of two smaller than 
the corresponding parton gas yield at LHC.

Our results show that, regardless
of the initial conditions, the
massive excitations of the quarks and gluons significantly enhance charm
production by the plasma.  Introducing thermal gluon and quark masses slowed
the expansion of the system.   The longer lifetime of the plasma as well as the
reduced threshold for charm production with massive quarks and gluons leads to
an enhancement of charm production over the massless case.
Thus charm enhancement
in heavy ion collisions could be an excellent probe of
the presence of collective excitations in the deconfined plasma. If such 
enhanced charm production, beyond that predicted
from the initial nucleon-nucleon collisions is observed, it could be expected
that other exotic phenomena due to massive quark and gluon excitations 
may be found.

\section*{Acknowledgments}

We would like to thank D. Rischke for enlightening discussions on the equation
of state.
The authors thank the Institute for Nuclear Theory at the 
University of Washington in Seattle for their hospitality.
This work was supported in part by the Director,
Office of Energy Research, Division of Nuclear Physics of the Office of High
Energy and Nuclear Physics of the U. S. Department of Energy under Contract
Number DE-AC03-76SF0098, 
by the National Scientific Research Fund (Hungary), OTKA 
No. T014213 and F019689 as well as 
by the U.S.-Hungarian Science and Technology Joint Fund, No. 378/93
and the Foundation for the Hungarian Science (MHB).

\vfill

\newpage

\begin{table}
\begin{center}
\begin{tabular}{|c|c|c|c|c|} \hline
\multicolumn{5}{|c|}{RHIC} \\ \hline
 & \multicolumn{2}{c|}{$m_c = 1.2$ GeV} & \multicolumn{2}{c|}{$m_c = 1.5$
 GeV} \\ \hline
 & parton gas & minijet gas & parton gas & minijet gas \\ \hline
$m=0$ $g$ & 3.2  & 0.053 & 0.93 & 0.0128 \\ \hline
$m \neq 0$ $g$ & 4.6 & 0.065 & 1.27 & 0.0154 \\ \hline
$m=0$ $g+q$ & 3.7 & 0.014 & 1.07 & 0.0027 \\ \hline
$m \neq 0$ $g+q$ & 4.9 & 0.016 & 1.38  & 0.0030 \\ \hline
\end{tabular}
\caption[]{The total thermal $c \overline c$ pair yield from a parton gas and
a minijet gas at RHIC.  We consider both a massless and massive pure
gluon gas and a quark-gluon system with massless and massive components.}
\end{center}
\end{table}
\vfill
\clearpage

\begin{table}
\begin{center}
\begin{tabular}{|c|c|c|c|c|} \hline
\multicolumn{5}{|c|}{LHC} \\ \hline
 & \multicolumn{2}{c|}{$m_c = 1.2$ GeV} & \multicolumn{2}{c|}{$m_c = 1.5$
 GeV} \\ \hline
 & parton gas & minijet gas & parton gas & minijet gas \\ \hline
$m=0$ $g$ & 102 & 39 & 43 & 21 \\ \hline
$m \neq 0$ $g$ & 198 & 101 & 76 & 47 \\ \hline
$m=0$ $g+q$ & 145 & 22 & 60 & 9.7 \\ \hline
$m \neq 0$ $g+q$ & 245 & 38 & 94 & 15.6 \\ \hline
\end{tabular}
\caption[]{The total thermal $c \overline c$ pair yield from a parton gas and
a minijet gas at LHC.  We consider both a massless and massive pure
gluon gas and a quark-gluon system with massless and massive components.}
\end{center}
\end{table}
\vfill
\clearpage

\noindent {\Large\bf Figure Captions}

\begin{description}

\item[Fig. 1:] The time
evolution of the plasma produced in Au+Au collisions at RHIC
is examined for a parton gas in (a), (c), (e) and (g) and a minijet gas in
(b), (d), (f) and (h).  The temperature is shown in (a) and (b), the energy
density in (c) and (d), the square of the sound speed in (e) and (f) and the
plasma volume in (g) and (h).  The solid curve is for a
massless gluon gas only while the dashed curve is for a gas of massive gluons.
The dot-dashed curve is calculated assuming massless gluons and light quarks
and $m_s=150$ MeV for the strange quark.  The dotted curve is the result when
the quarks and gluons have effective masses.

\item[Fig. 2:] The same as fig.\ 1 for Pb+Pb collisions at LHC.

\item[Fig. 3:]  Time dependence of the effective quark and gluon masses.  
The parton gas results are given in (a) RHIC and (b) LHC.  The minijet gas
results are shown in (c) RHIC and (d) LHC.  The RHIC results are given for 
Au+Au
collisions, the LHC results for Pb+Pb collisions.  The solid curve is the
effective light quark mass, the dashed is the strange quark mass.  The
dot-dashed curve is the effective gluon mass.

\item[Fig. 4:] The production rate (a), (c) and $c \overline c$ pair yield
(b), (d) are given for Au+Au collisions at RHIC with $m_c = 1.2$ GeV.  The 
parton gas results are given in (a) and (b) while the minijet gas results are
shown in (c) and (d).   The solid curve is for a
massless gluon gas only while the dashed curve is for a gas of massive gluons.
The dot-dashed curve is calculated assuming massless gluons and light quarks
and $m_s=150$ MeV for the strange quark.  The dotted curve is the result when
the quarks and gluons have effective masses.

\item[Fig. 5:] The same as in fig.\ 4 for Au+Au collisions at RHIC with
$m_c = 1.5$ GeV.

\item[Fig. 6:] The same as in fig.\ 4 for Pb+Pb collisions at LHC with
$m_c = 1.2$ GeV.

\item[Fig. 7:] The same as in fig.\ 4 for Pb+Pb collisions at LHC with
$m_c = 1.5$ GeV.

\item[Fig. 8:] The $c \overline c$ pair mass distribution from Au+Au collisions
at RHIC with $m_c = 1.2$ GeV (a), (b) and $m_c = 1.5$ GeV (c), (d).  The
parton gas results are given in (a) and (c) while the minijet gas results are
shown in (b) and (d).  The solid curve is for a
massless gluon gas only while the dashed curve is for a gas of massive gluons.
The dot-dashed curve is calculated assuming massless gluons and light quarks
and $m_s=150$ MeV for the strange quark.  The dotted curve is the result when
the quarks and gluons have effective masses.

\item[Fig. 9:] The same as in fig.\ 8 for Pb+Pb collisions at LHC.

\end{description}
\vfill
\eject

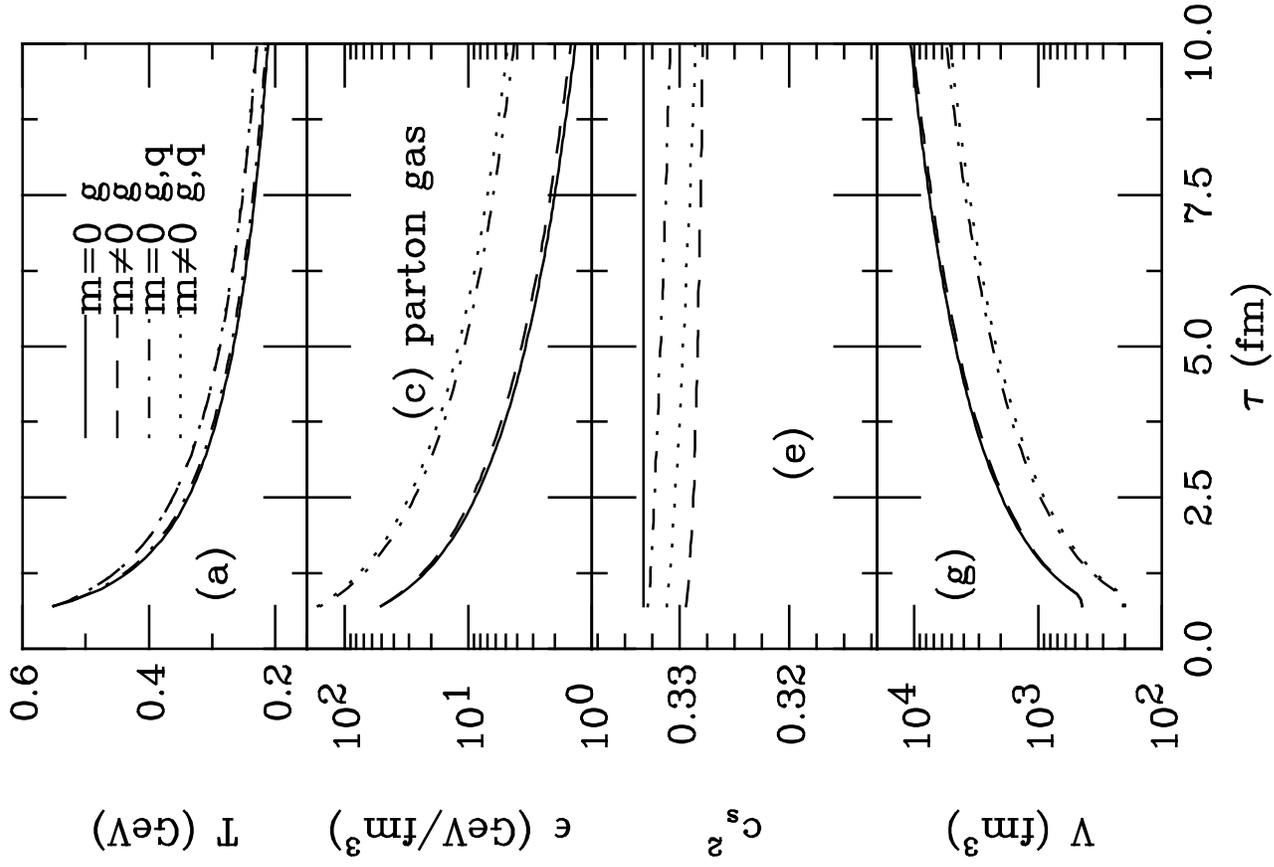
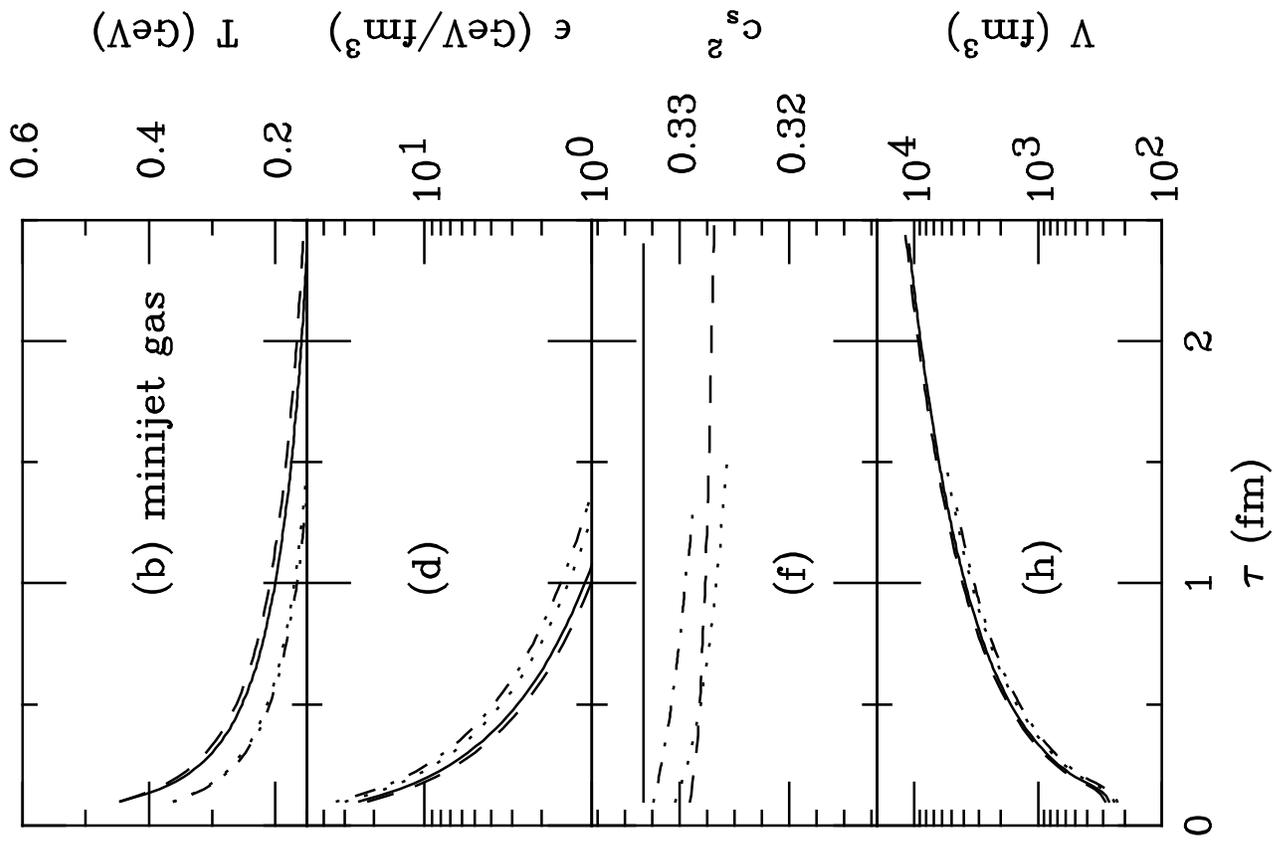
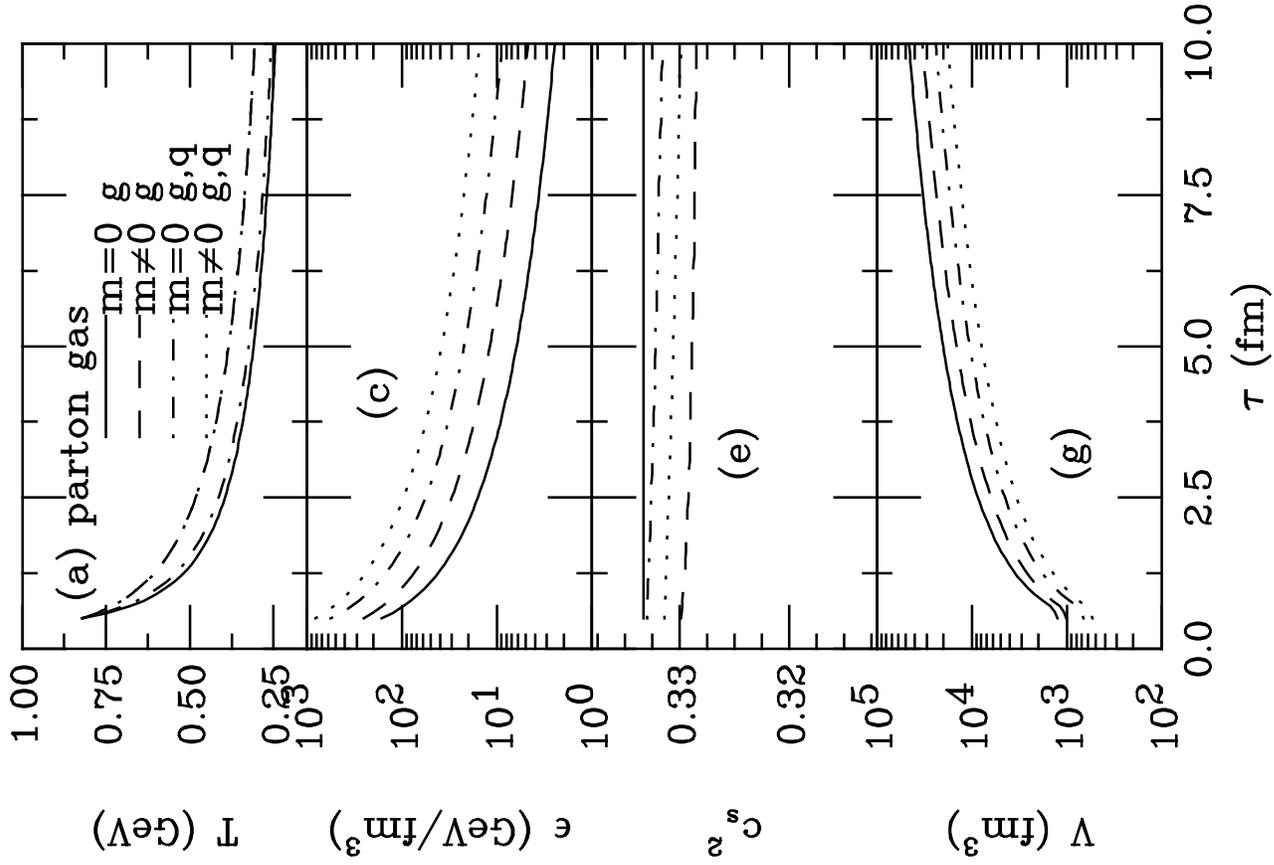
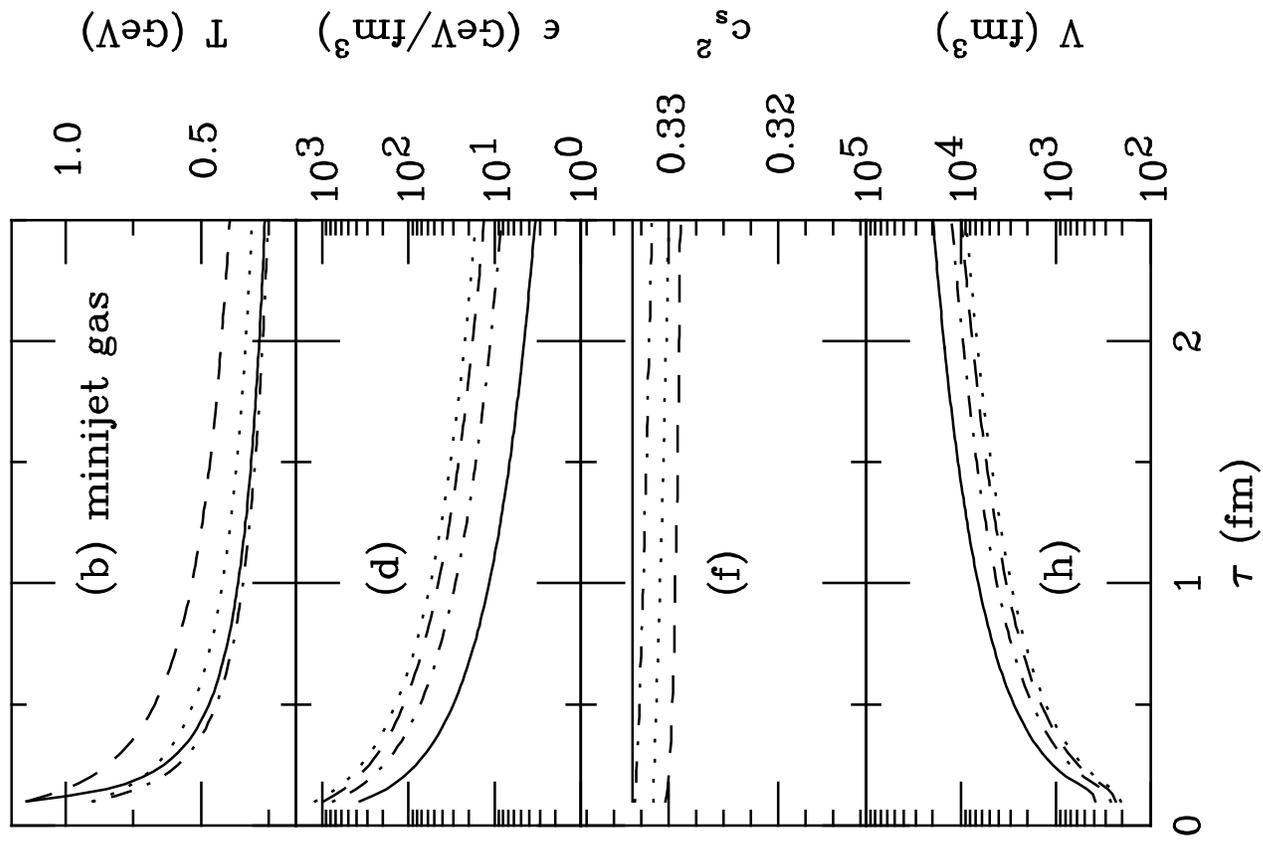
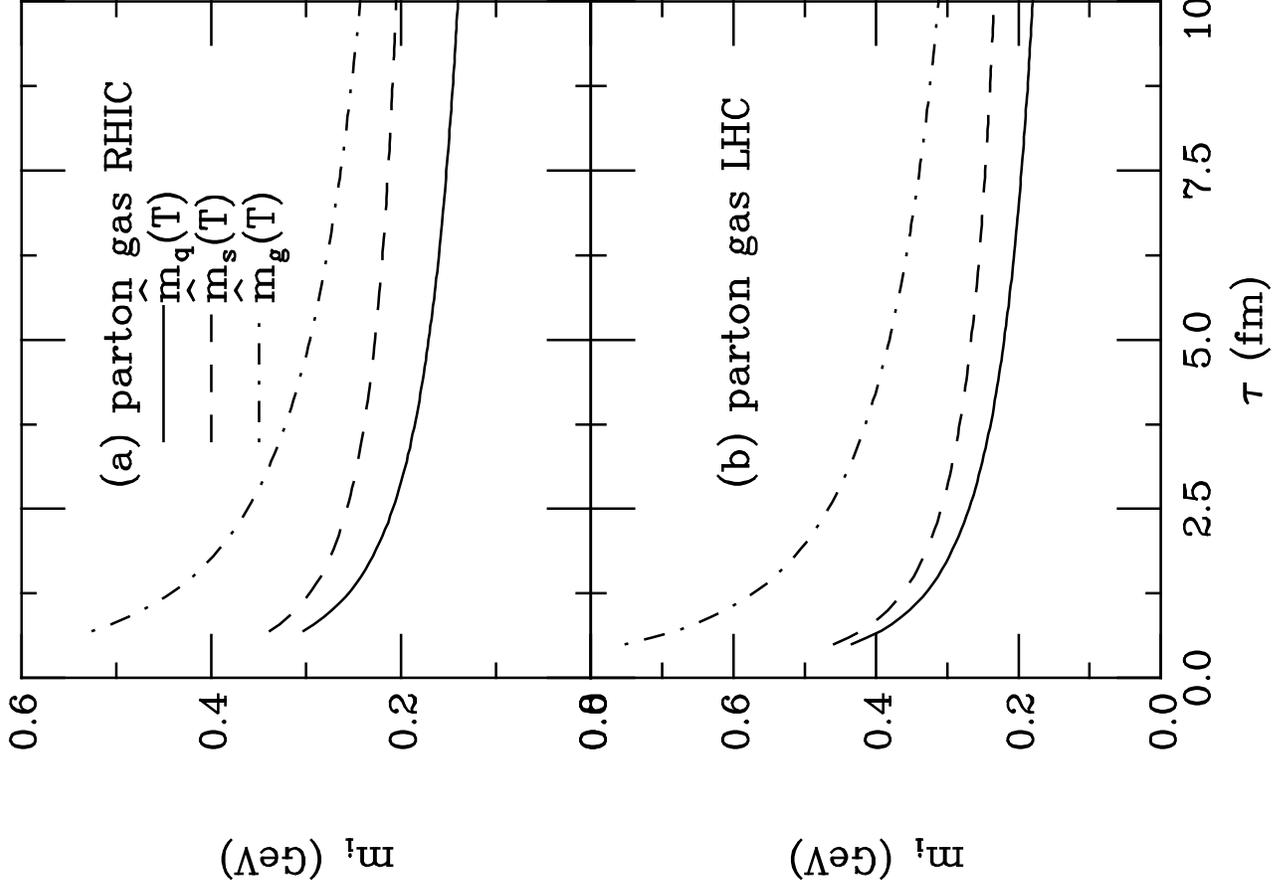
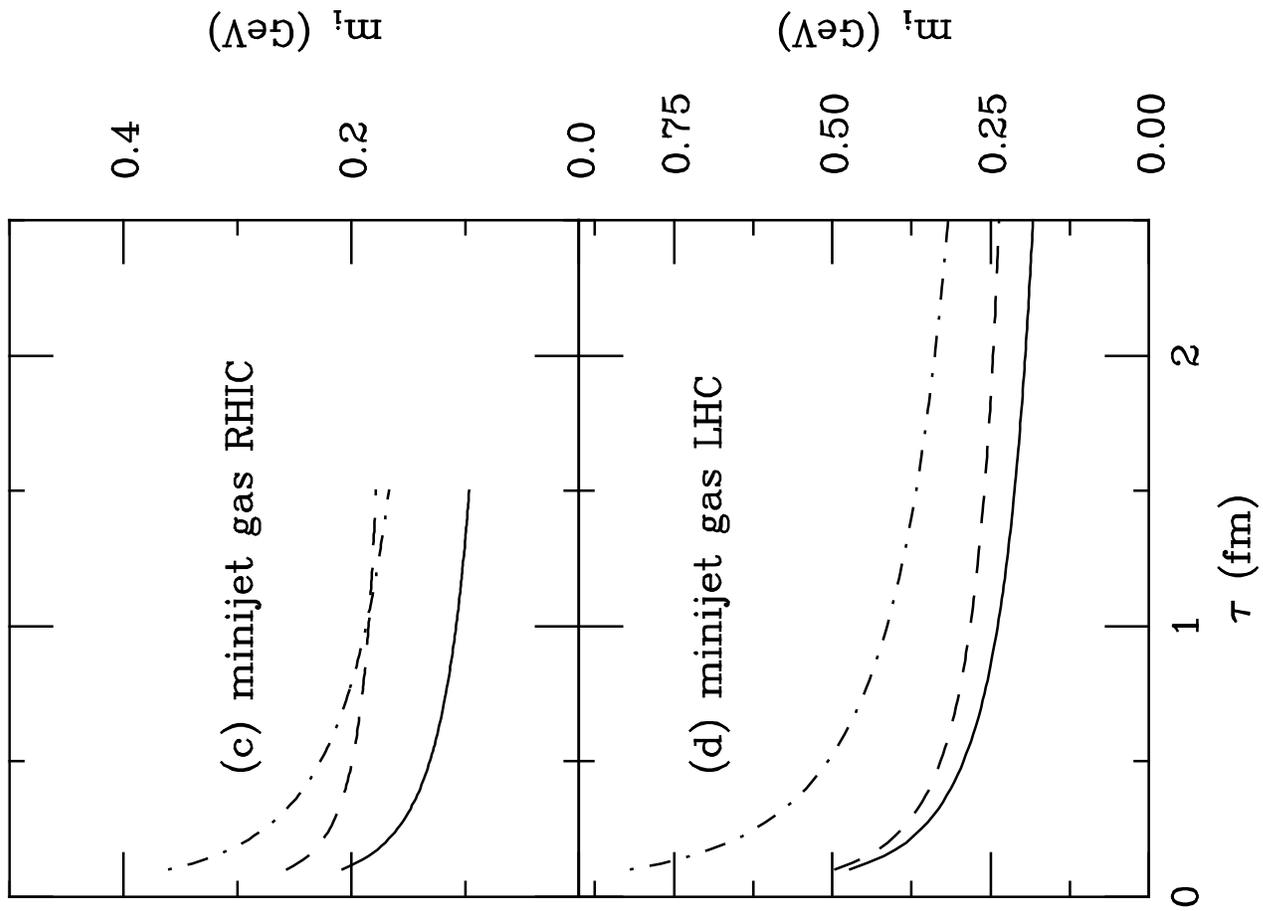
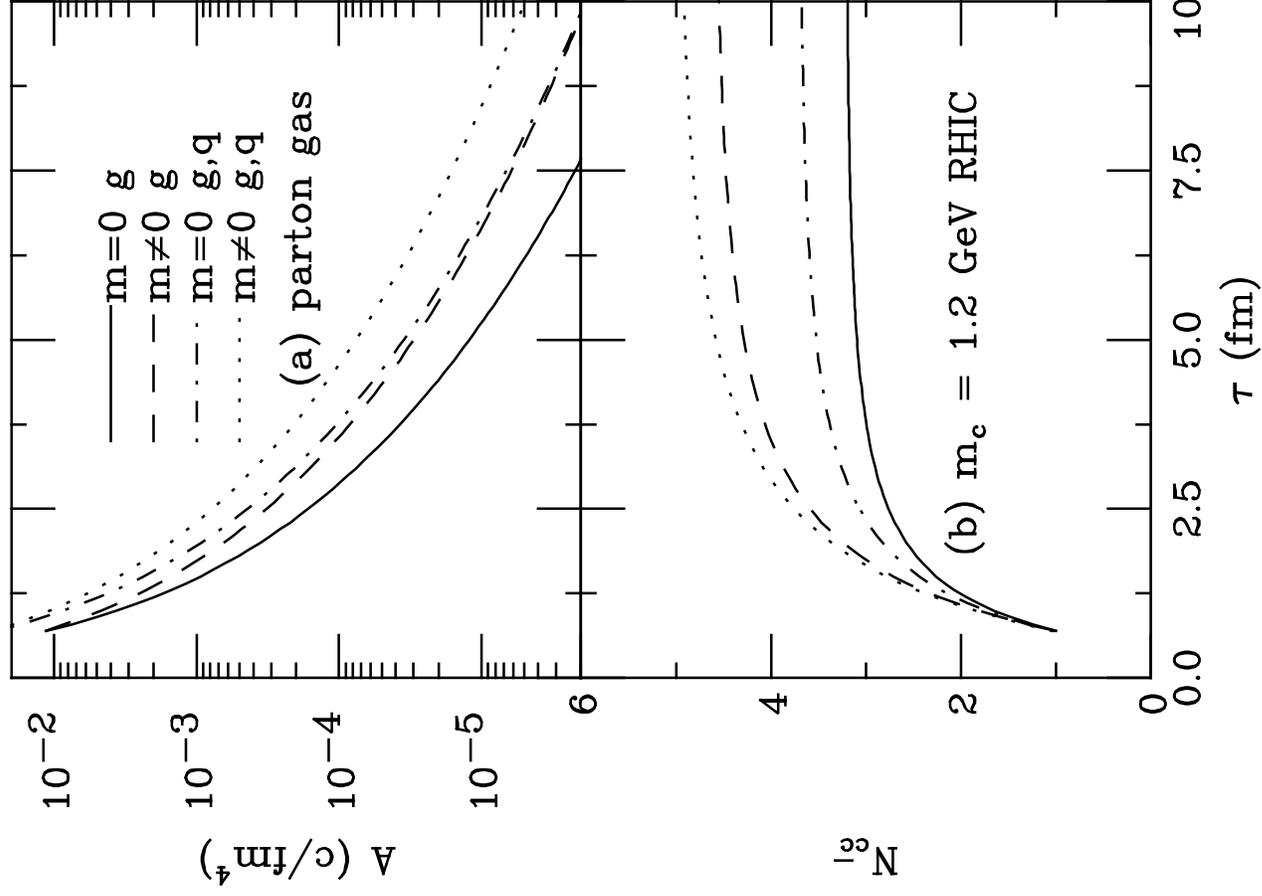
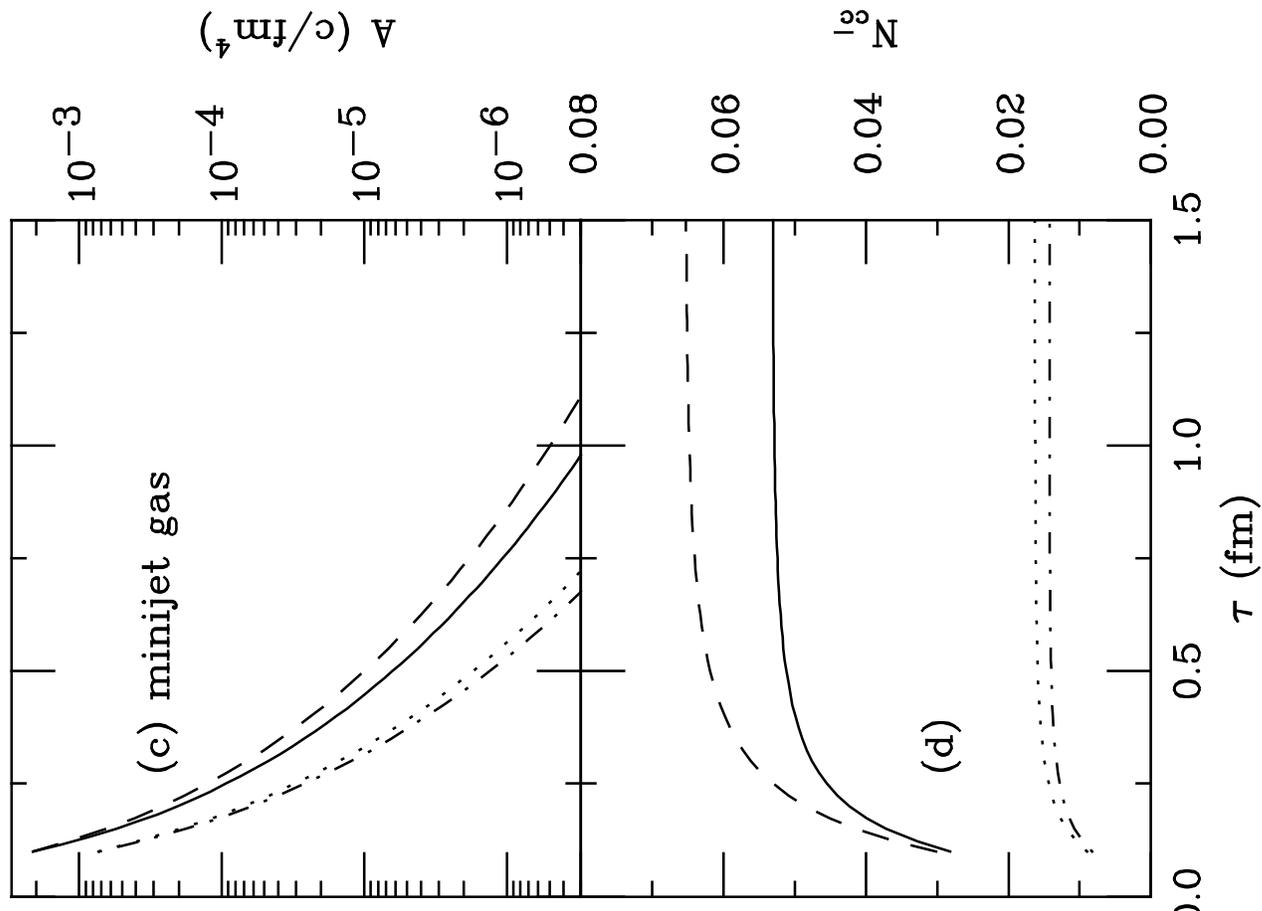
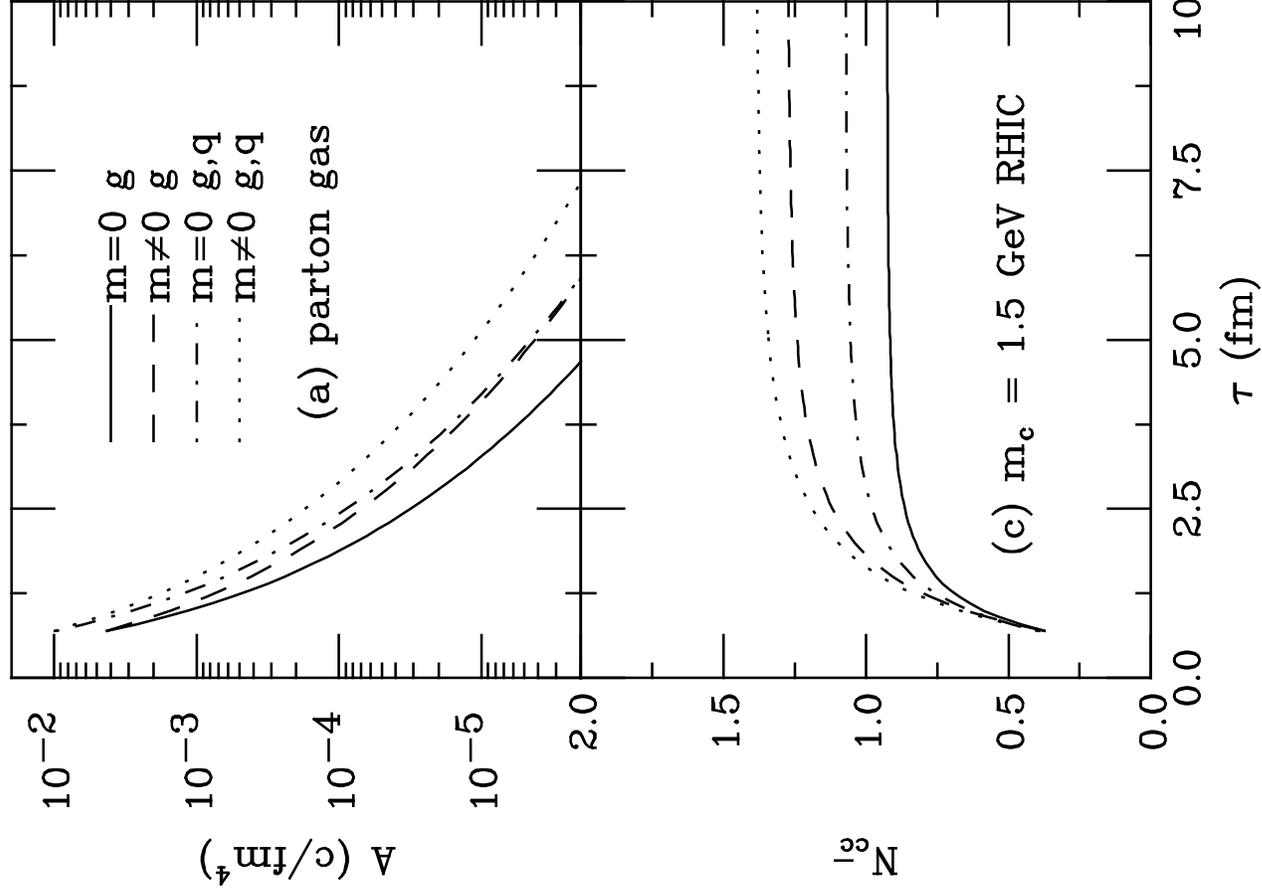
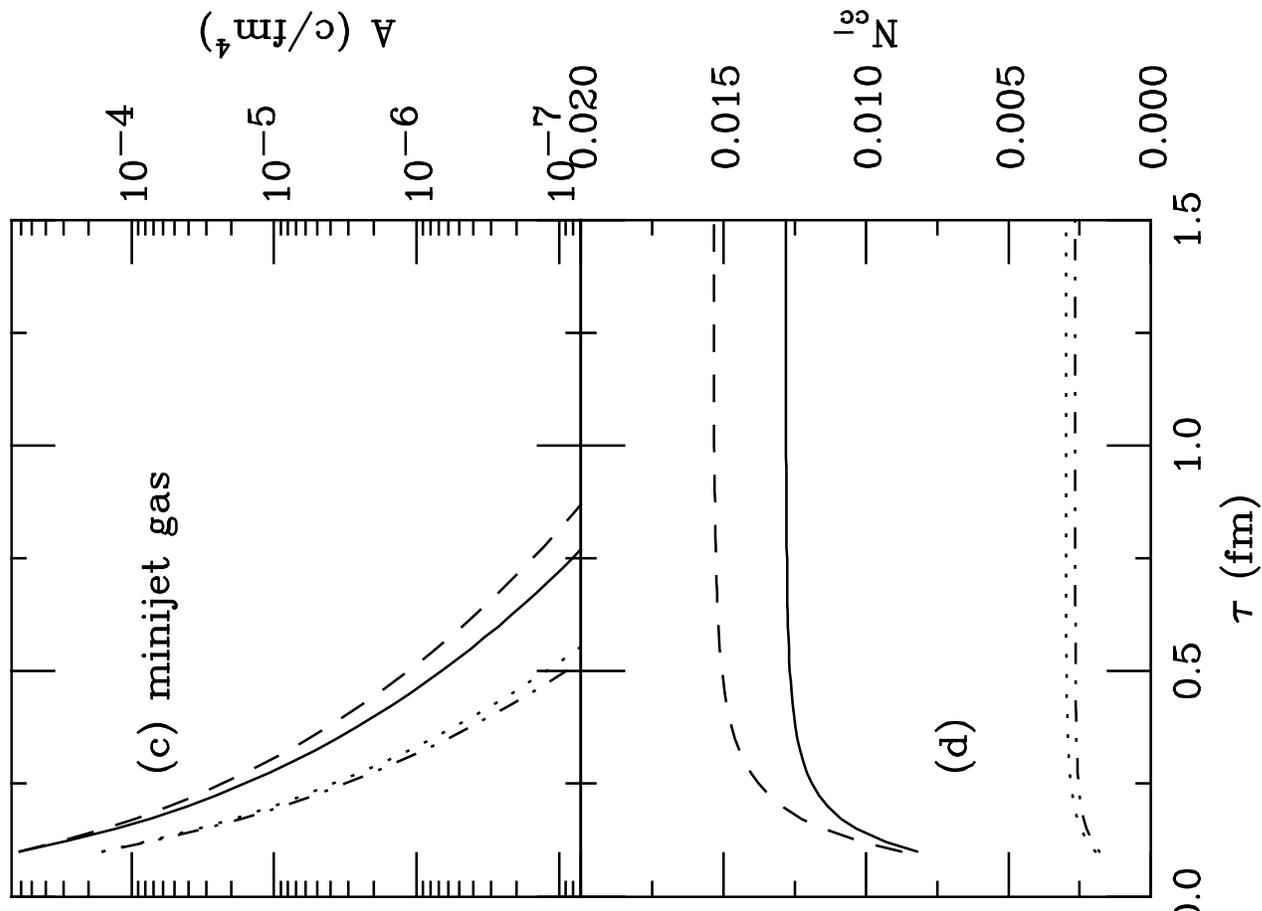
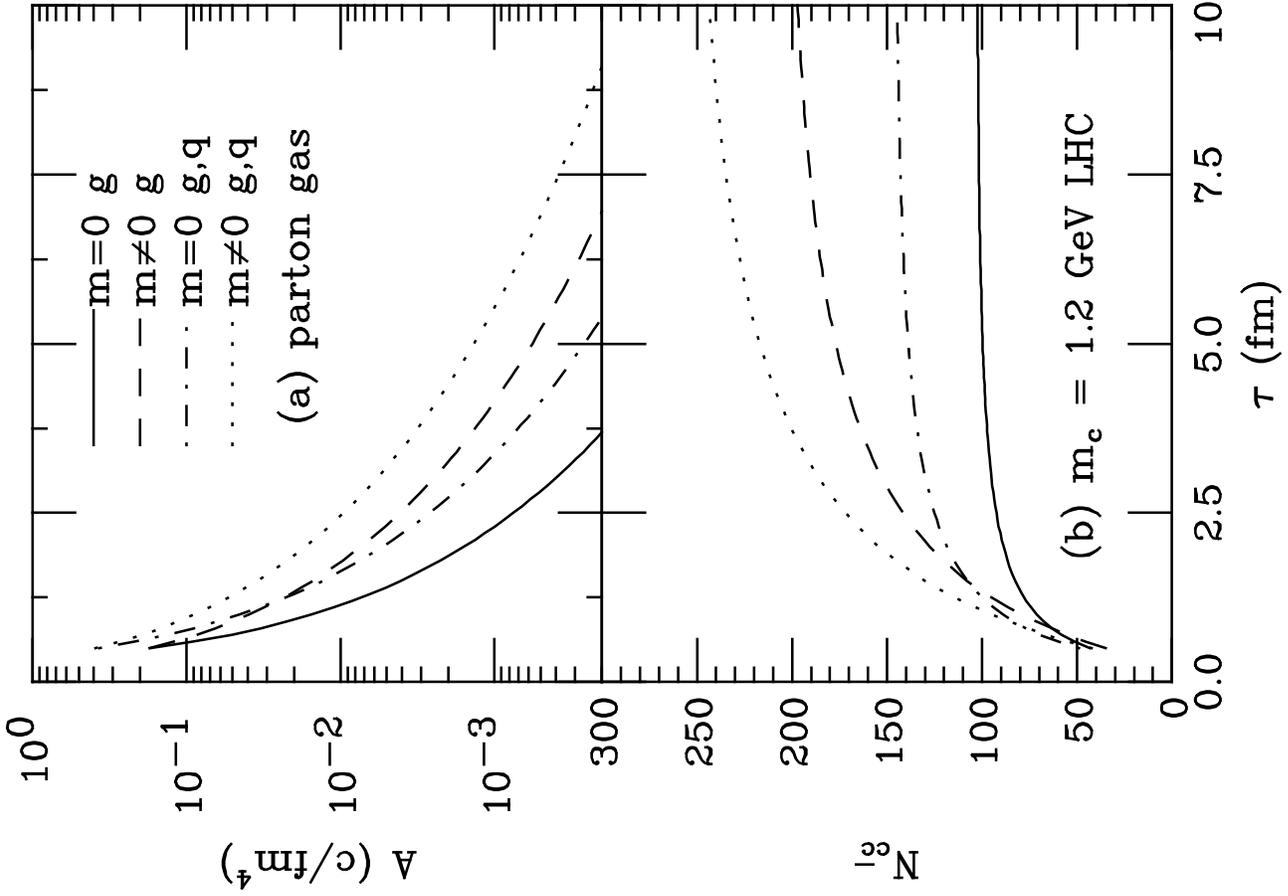
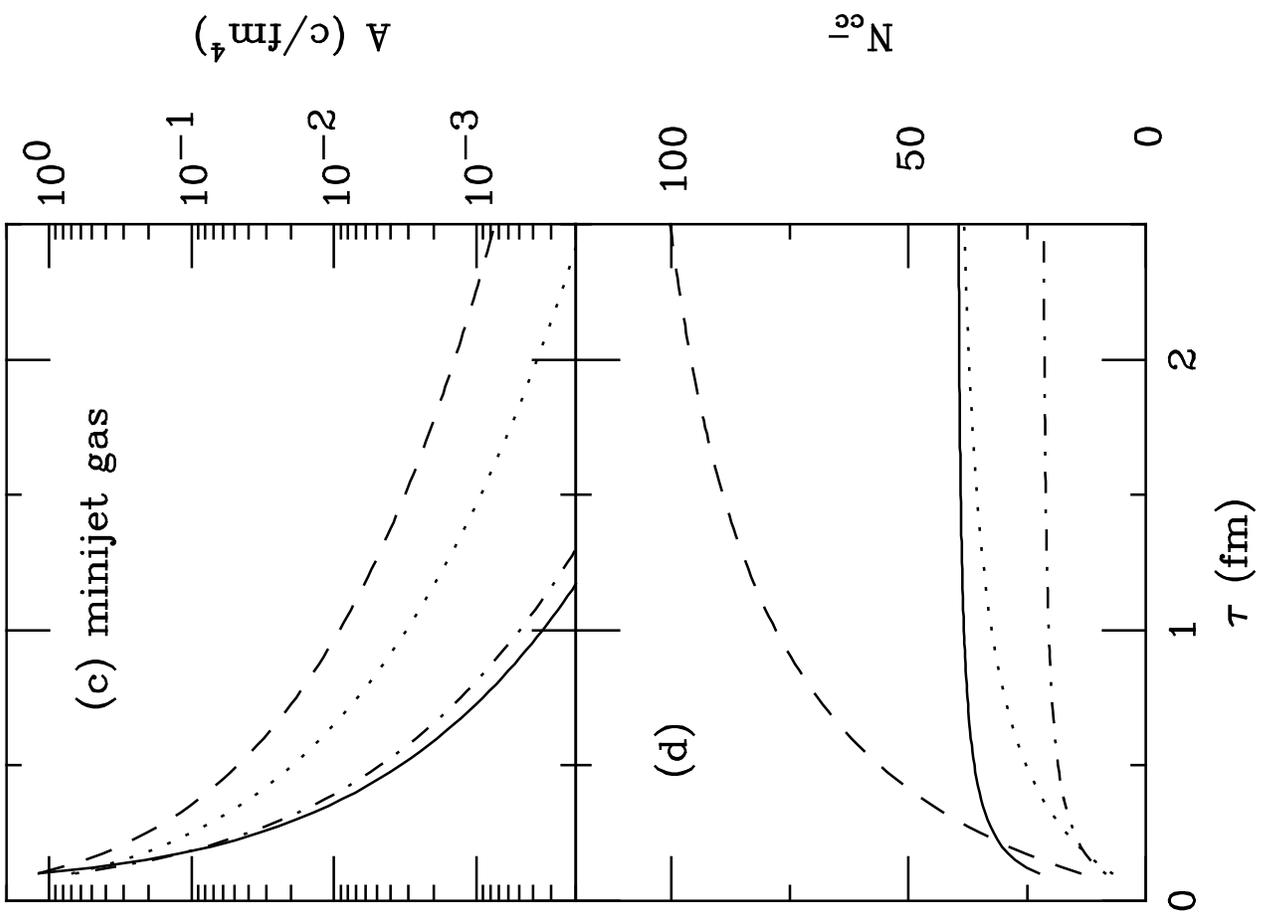
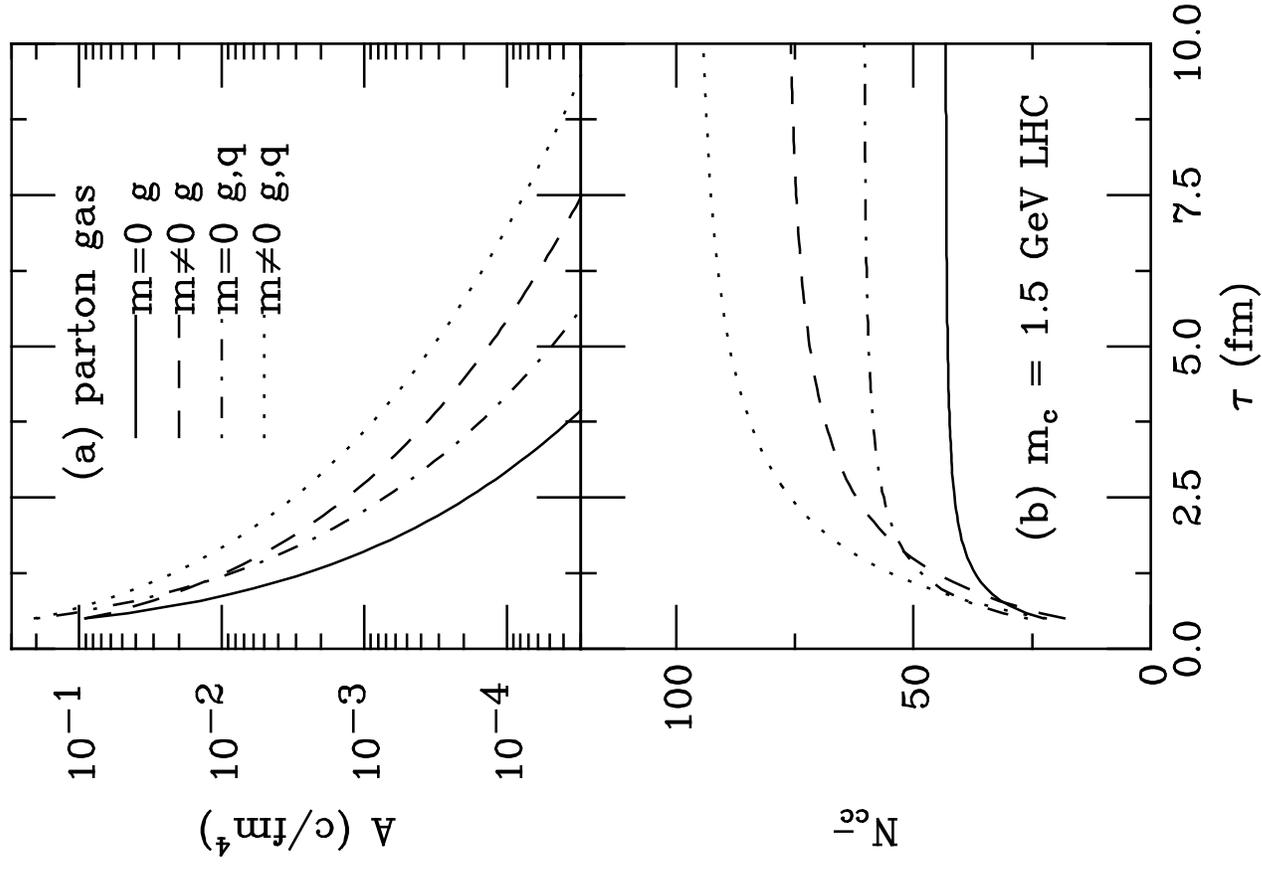
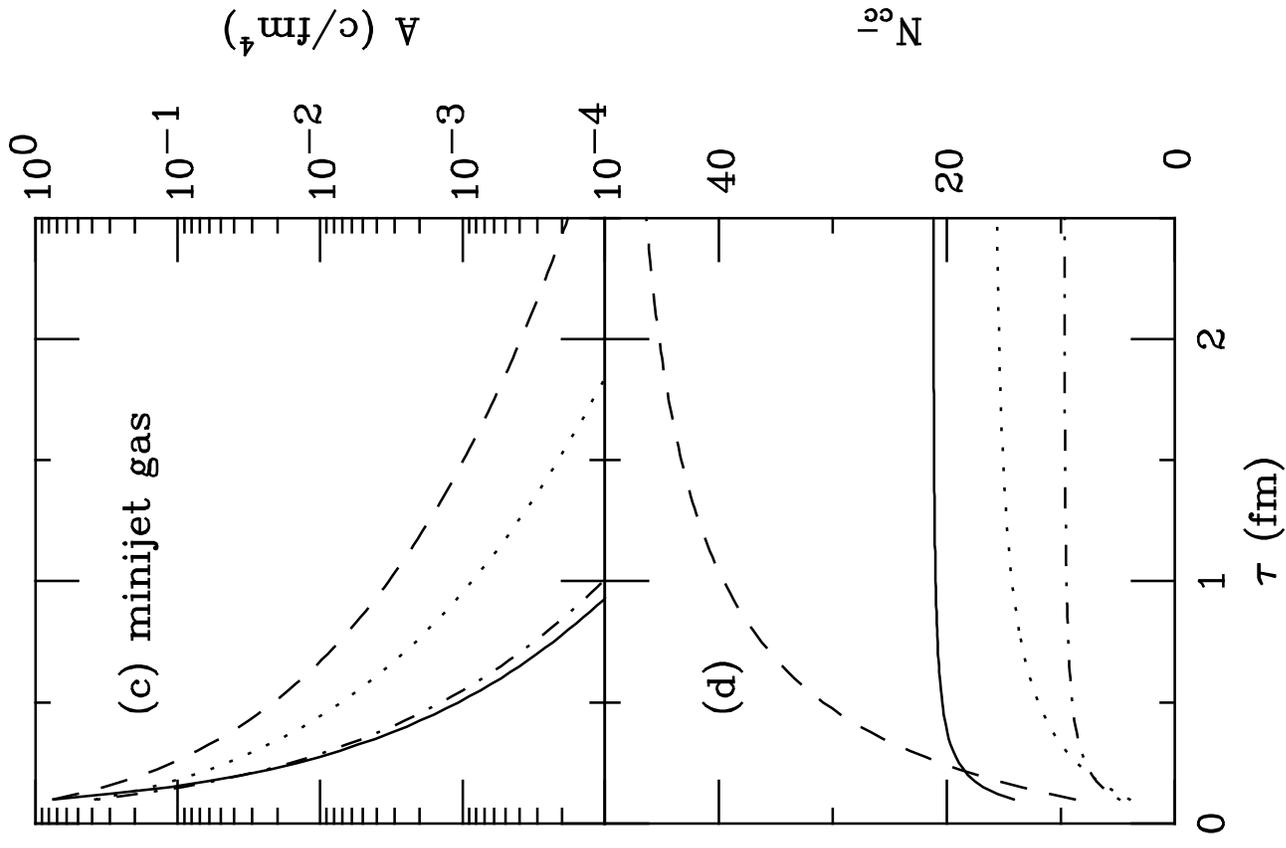
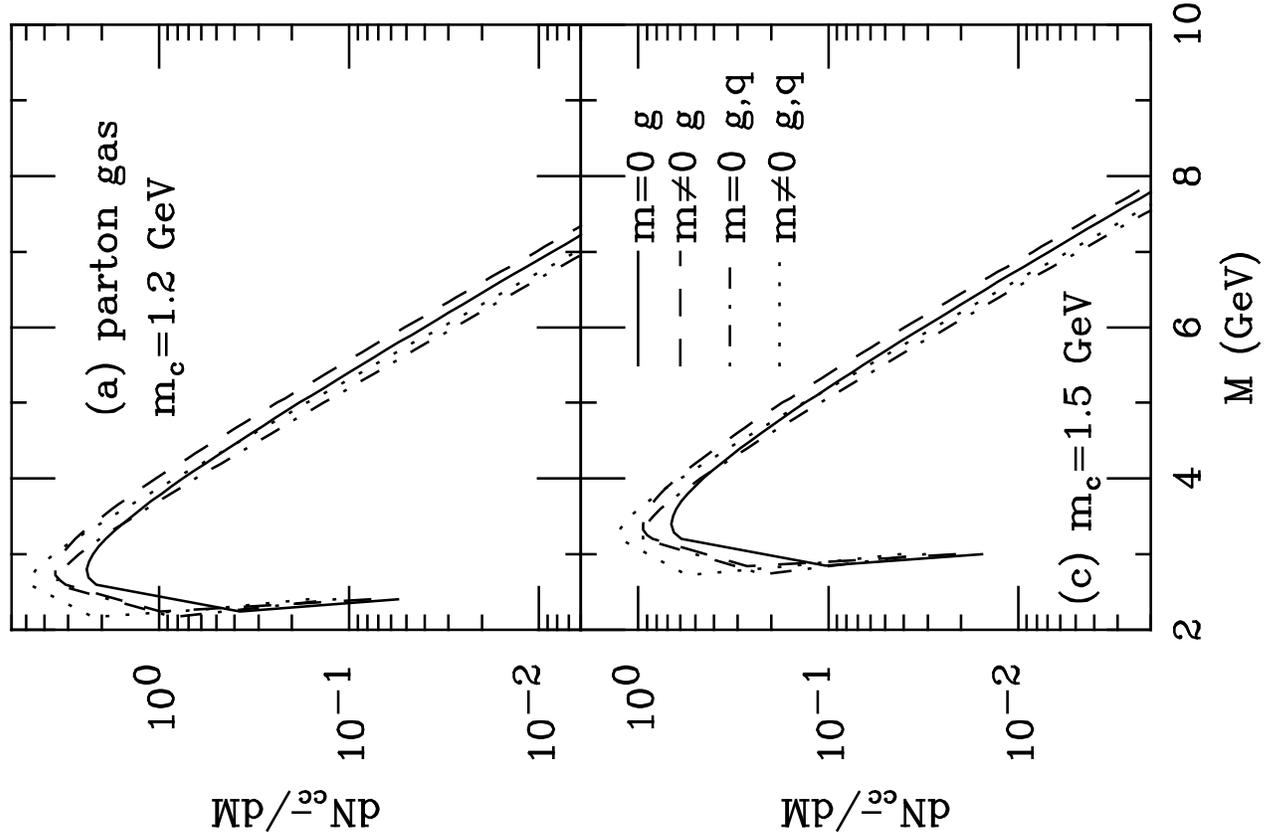
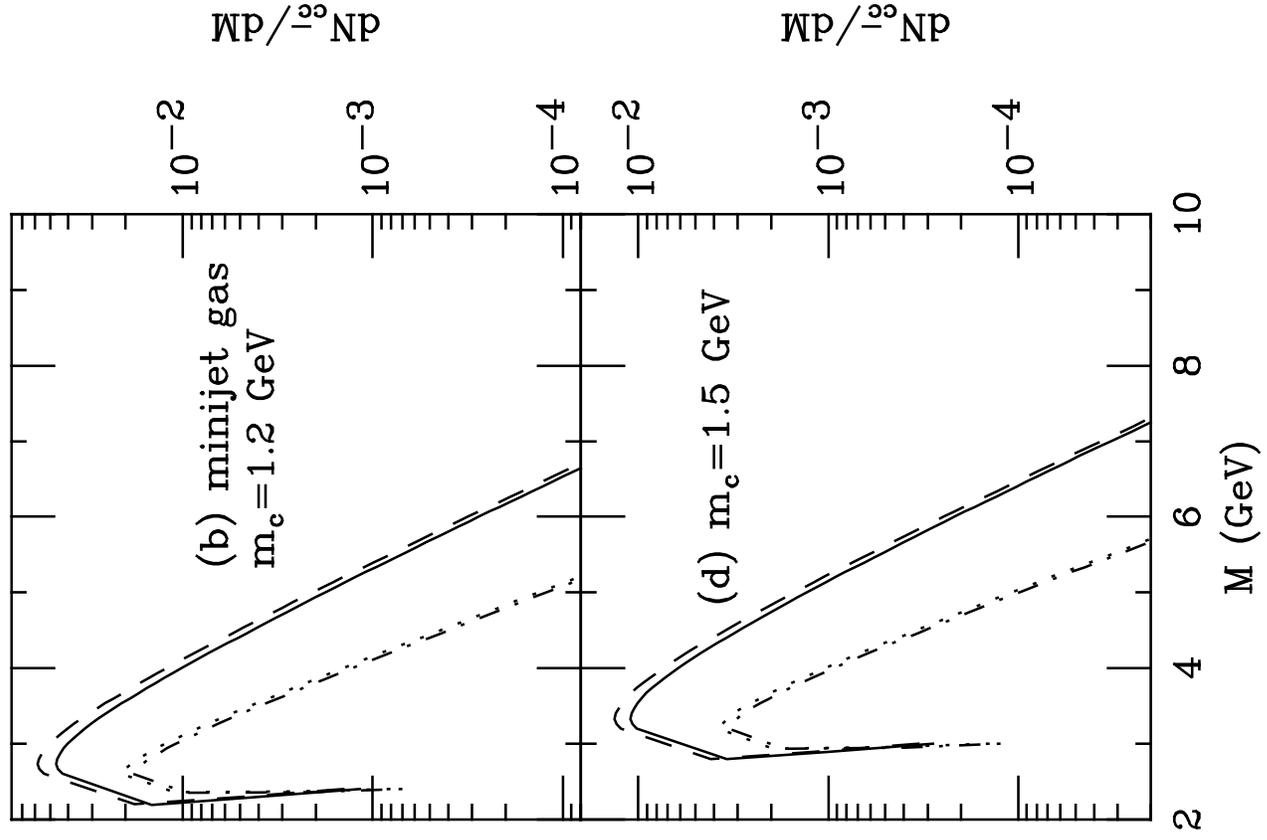
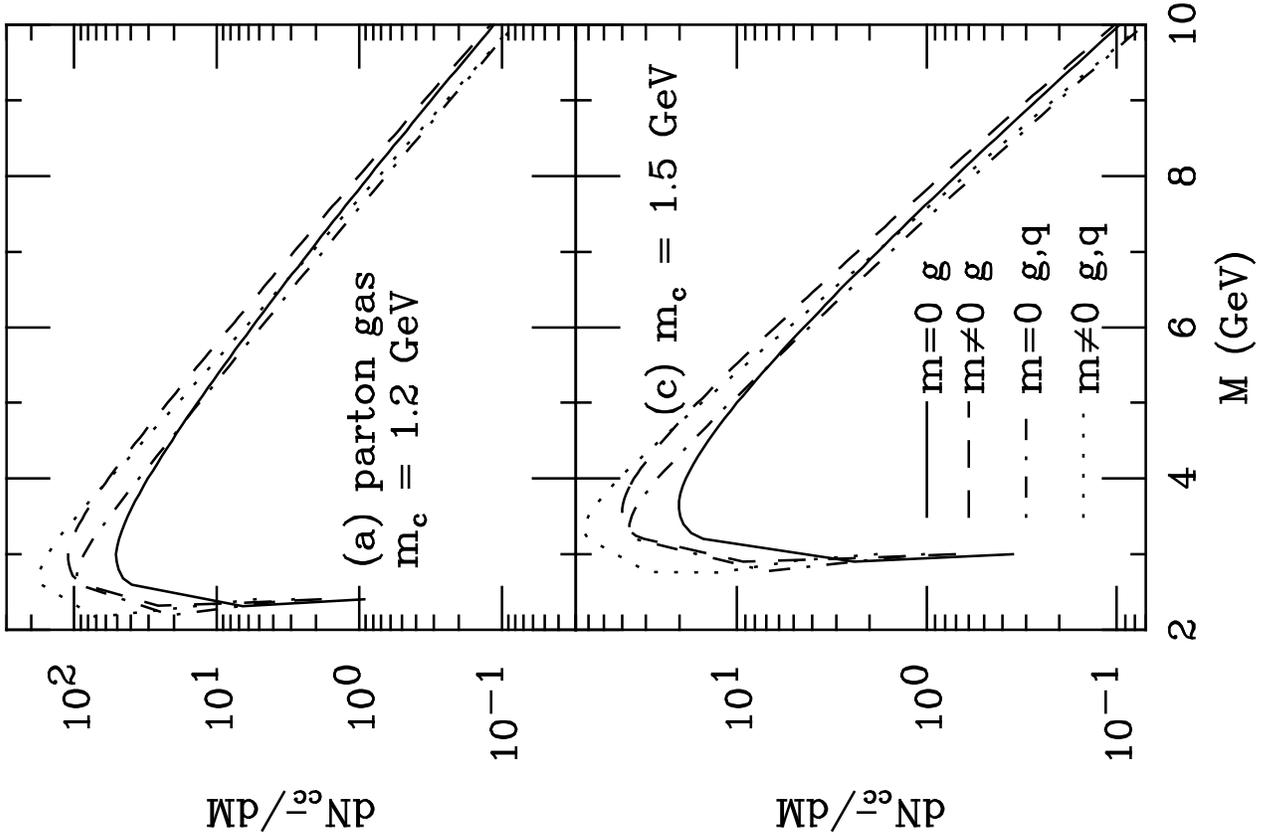
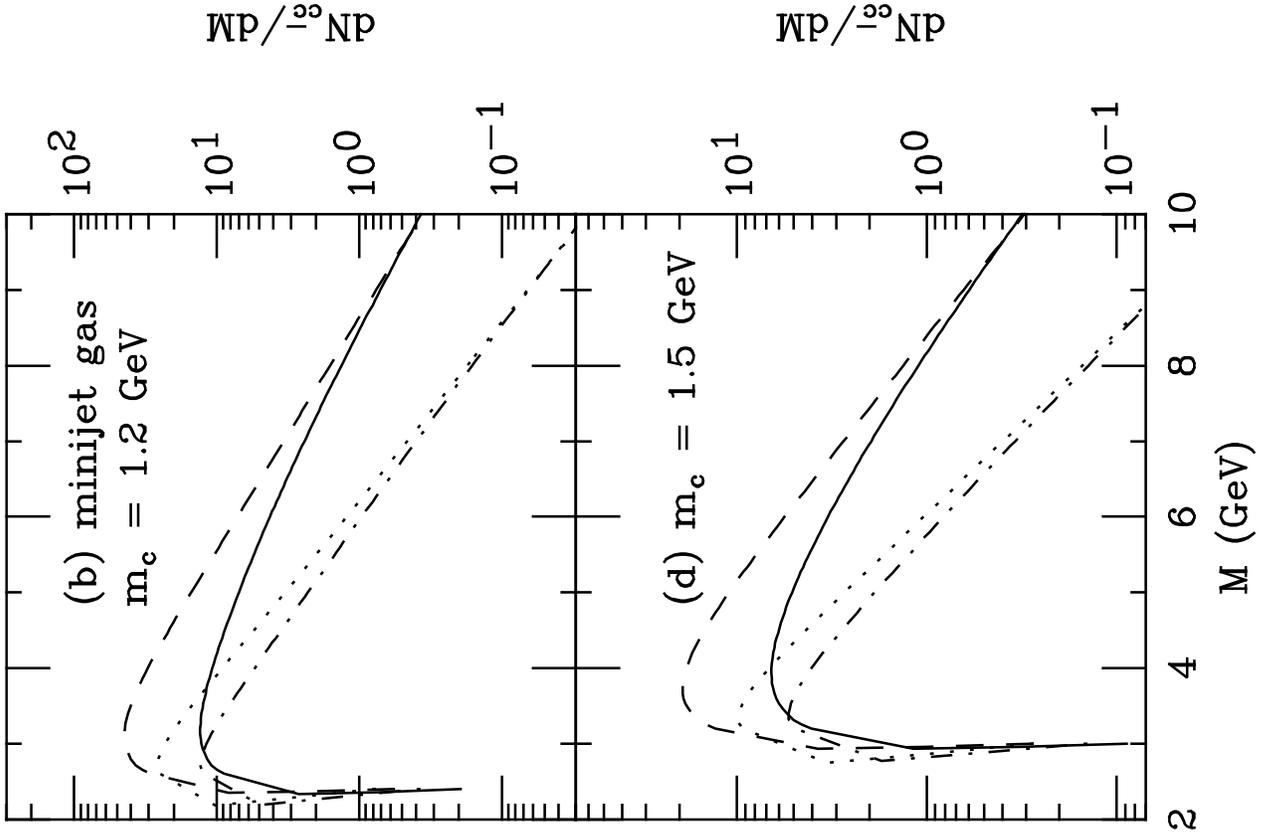
\end{document}